\pdfoutput=1
\documentclass[cits]{JINST}
\usepackage{graphicx}
\usepackage{textcomp} 
\usepackage{amsmath,amsfonts,amssymb,mathrsfs}
\def\degC{$^{\circ}\mathrm{C}$\;}

\def\degC{${}^{\circ}\mathrm{C}$\;}
\def\microns{$\rm{{\mu}m}$\;}

\title{The design, construction and performance of the MICE target}

\author{C.N.~Booth\thanks{Corresponding author}, P.~Hodgson, L.~Howlett, R.~Nicholson, E.~Overton, M.~Robinson, P.J.~Smith\\
   Department of Physics and Astronomy, University of Sheffield, Sheffield S3 7RH, UK
   \email{C.Booth@sheffield.ac.uk}
  } 
\author{M.~Apollonio$^a$, G.~Barber, A.~Dobbs, J.~Leaver, K.R.~Long\\
  Department of Physics, Blackett Laboratory, Imperial College
    London, Exhibition Road, London SW7 2AZ, UK\\
    \llap{$^a$}Now at Diamond Light Source Ltd
  } 
\author{B.~Shepherd\\
    STFC Daresbury Laboratory, Daresbury, Cheshire WA4 4AD, UK
  } 
\author{D.~Adams, E.~Capocci, E.~McCarron, J.~Tarrant\\
  STFC Rutherford Appleton Laboratory, Chilton, Didcot, 
    Oxfordshire, OX11 0QX, UK
  } 
\abstract{
The pion-production target that serves the MICE Muon Beam consists of
a titanium cylinder that is dipped into the halo of the ISIS proton
beam.
The design and construction of the MICE target system are described
along with the quality-assurance procedures, electromagnetic drive
and control systems, the readout electronics, and the data-acquisition
system. 
The performance of the target is presented together with the particle
rates delivered to the MICE Muon Beam.
Finally, the beam loss in ISIS generated by the operation of the target
is evaluated as a function of the particle rate, and the operating
parameters of the target are derived.
}

\keywords{Accelerator applications; Targets;
Instrumentation for particle accelerators and storage rings;
Control and monitor systems online;
Overall mechanics designs}

\preprint{MICE-PUB-BEAM-392\\IC/HEP/12-07\\RAL-P-2012-007}

\begin{document}
%
%
%
%
%
\section{Introduction}
\label{Sect:Intro}

Muon storage rings have been proposed for use as sources of intense
high-energy neutrino beams in a Neutrino Factory \cite{Geer:1997iz}
and as the basis for multi-TeV lepton-antilepton colliding-beam
facilities \cite{MC}.
To optimise the performance of such facilities requires the
phase-space compression (cooling) of the muon beam prior to
acceleration and storage.  
The short muon-lifetime makes it impossible to employ traditional
techniques to cool the beam while maintaining the muon-beam intensity. 
Ionisation cooling, a process in which the muon beam is
passed through a series of liquid-hydrogen absorbers interspersed with
accelerating RF cavities, is the technique proposed to cool the beam. 
The international Muon Ionisation Cooling Experiment (MICE) 
will provide an engineering demonstration of the ionisation-cooling
technique and will allow the factors affecting the performance of
ionisation-cooling channels to be investigated in detail \cite{MICE}.
Muon beams of momenta between 140\,MeV/c and 240\,MeV/c, with
normalised emittances between 2\,$\pi$mm and 10\,$\pi$mm, will be
provided by a purpose-built beam line on the 800\,MeV proton
synchrotron, ISIS \cite{ISIS}, at the Rutherford Appleton Laboratory
\cite{RAL}.

\begin{figure}
  \begin{center}
    \includegraphics[width=0.95\textwidth,bb=0 0 800 272]
      {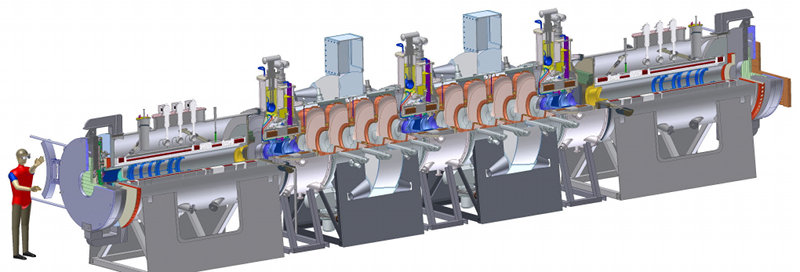}
  \end{center}
  \caption{
    Cutaway 3D rendering of the international Muon Ionisation 
    Cooling Experiment (MICE).
    The muon beam enters from the bottom left of the figure.
    The upstream PID instrumentation (not shown) is composed of two 
    time-of-flight hodoscopes (TOF0 and TOF1) and two threshold Cherenkov
    counters (CKOVa and CKOVb). The upstream spectrometer is followed by
    the MICE cooling channel, which is composed of three 20\,$l$ volumes of
    liquid hydrogen and two sets of four 201\,MHz accelerating cavities
    embedded in a solenoidal transport channel. This in turn is followed by
    the downstream spectrometer, a third time-of-flight hodoscope (TOF2), 
    and a calorimeter system (KL and EMR).
  }
  \label{Fig:Intro:MICE}
\end{figure}

MICE is a single-particle experiment in which the position and
momentum of each muon is measured before it enters the MICE cooling
channel and once again after it has left (see figure
\ref{Fig:Intro:MICE}) \cite{TRD}.
The MICE cooling channel, which is based on one lattice cell of the
cooling channel described in \cite{StudyII}, comprises three 20\,$l$
volumes of liquid hydrogen and two sets of four 201\,MHz accelerating
cavities.
Beam transport is achieved by means of a series of superconducting
solenoids.
A particle-identification (PID) system (scintillator time-of-flight
hodoscopes TOF0 and TOF1 and threshold Cherenkov counters CKOVa and
CKOVb) upstream of the cooling channel allows a pure muon beam to be
selected. 
Downstream of the cooling channel, a final hodoscope (TOF2) and a
calorimeter system allow muon decays to be identified.  
The calorimeter is composed of a KLOE-like lead-scintillator section (KL)
followed by a fully active scintillator detector (the electron-muon ranger, 
EMR) in which the muons are brought to rest. 
For a full description of the experiment see \cite{TRD}.

\begin{figure}
  \begin{center}
    \includegraphics[height=8cm]{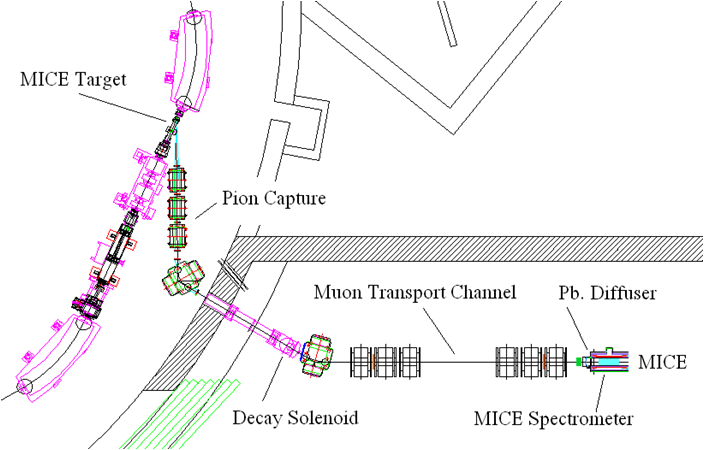}
  \end{center}
  \caption{
    The MICE Muon beam-line
  }
  \label{Fig:Intro:MMB}
\end{figure}

A schematic diagram of the MICE Muon Beam is shown in figure
\ref{Fig:Intro:MMB} \cite{MICE_Beam}.
A cylindrical target is dipped into the edge of the circulating proton
beam.
The depth at which the target is dipped into the proton beam is
characterised by the `beam centre distance' (BCD) which is defined to
be the distance from the tip of the target to the proton-beam axis at
the target's maximum excursion into the beam.  
Pions produced in the target are captured by a quadrupole triplet and
transported to a dipole magnet by which the pion momentum is selected.
A 5\,T super-conducting `decay' solenoid follows the dipole.  
The additional pion path-length in the decay solenoid increases the
muon-production efficiency.  
Following the solenoid, a second dipole is used to select the muon
momentum and the beam is transported to MICE using a pair of
large-aperture quadrupole triplets. 

This paper is organised as follows. 
The requirements for the target system and an overview of its design
are presented in section \ref{Sect:Requirements}. 
Section \ref{Sect:LinearMotor} describes the design of the linear
motor.
The mechanical design of the target mechanism and the mechanical
interface to the ISIS accelerator is presented in
section \ref{Sect:Mechanical}.
Section \ref{Sect:OpticalPosition} describes the optical position-measurement 
system, the power electronics used to drive the linear motor is described in 
section \ref{Sect:PowerElectronics} and the control system in
section \ref{Sect:Controls}.
The performance of the system, the particle rate delivered
for the MICE Muon Beam and the beam losses induced in ISIS are
presented in section \ref{Sect:Performance}.
Finally, a summary is given in section \ref{Sect:Summary}.

%
\section{Requirements and overview}
\label{Sect:Requirements}
%
%
The ISIS synchrotron \cite{ISIS} operates on a basic cycle of 50\,Hz.
Protons are injected with a kinetic energy of 70\,MeV and accelerated 
to 800\,MeV over a period of 10\,ms prior to extraction.
In the following 10\,ms, the currents in the focusing and bending magnets are
reduced to their initial values, ready for the next pulse of protons to be
injected and accelerated.

MICE operation is parasitic to the functioning of ISIS and must cause 
minimal disruption to its principal function as a spallation neutron source.
On selected pulses, the MICE target is caused to dip into the outer low-density
halo of the proton beam just before extraction.
Pions produced in the target emerge through a thin window in the ISIS
vacuum system in the direction of the MICE muon beam line.
On injection, the proton beam effectively fills the beam pipe.
At the location of the MICE target, the beam has a vertical radius of $\sim 67$\,mm.
During acceleration, the beam shrinks to a radius of about 48\,mm.
To produce the required muon flux, the target must enter the beam by at least
5\,mm, so a minimum travel of 24\,mm is needed.
(In practice, the exact position of the edge of the beam and the intensity of 
the halo show long-term variations.
The position of the target at maximum insertion must therefore be 
controlled.)
The target must be outside the beam envelope for the first 8\,ms of the machine 
cycle, only entering the beam for the last 1 to 2\,ms before extraction, when
the protons are close to their maximum energy.
The exact time of insertion must also be controllable.

In order to meet the demands described above, a linear electromagnetic
drive was implemented to move the target vertically into the beam from
above.
The technical challenges are considerable.
The mechanism must be extremely reliable, to avoid disrupting normal
accelerator operation.
It must provide an acceleration of the order of 780\,ms$^{-2}$ so that the 
target overtakes the shrinking beam envelope and is removed before the 
next injection.
Operation must be precise and reproducible, both in position and timing 
relative to the beam cycle.
The mechanism must operate within a high radiation environment, and all
moving parts must use materials compatible with the stringent constraints 
of the accelerator's high-vacuum system. 
In case of any failure of the target mechanism, it must be possible to separate
it both mechanically and in terms of vacuum from the synchrotron.

The complete target mechanism, described in detail in this paper, consists of
a number of sub-assemblies.
\begin{itemize}
  \item The linear electromagnetic drive assembly, which contains:
    \begin{itemize}
      \item The shaft, forming the target at its lower end and carrying 
	a set of fixed permanent magnets, an optical readout vane 
	at its upper end and a stop to prevent the magnets falling from inside 
	the coils in the absence of power;
      \item A pair of bearings to support, guide and align the shaft;
      \item The stator, consisting of stationary coils with a water
	cooling system;
      \item A central steel tube forming a vacuum barrier between the target
	shaft and the stationary coil unit;
      \item An optical readout enclosure, with sapphire windows.
    \end{itemize}
  \item The mechanical support assembly, which contains:
    \begin{itemize}
     \item Flanges to provide accurate location of the stator and bearings;
     \item Conflat seals, to ensure the integrity of the vacuum system.
    \end{itemize}
  \item The mechanical and vacuum isolation system, to allow the unpowered
    target to be raised out of the beam, and the target vacuum to be 
    separated from the accelerator vacuum.
    This contains:
    \begin{itemize}
	\item A structural frame, carrying the weight of the target assembly;
	\item A jacking unit and support structure with motorised
	  screw-jack allowing a vertical travel of approximately 200\,mm;
	\item Centralising units and guide rods that guarantee the target
	  returns to its predefined position when lowered into its operating 
	  position;
	\item Edge-welded bellows, to allow relative movement of the components
	  under vacuum;
	\item A vacuum gate valve to isolate the vacuum systems.
    \end{itemize}
\end{itemize}
%
\section{Linear motor}
\label{Sect:LinearMotor}

%
%
%
A linear actuator was chosen as the most appropriate mechanism to drive the target
into the beam.
This implementation does not require any moving parts to cross the vacuum
chamber walls, and can be realised without the need for lubricated bearings.
For most of the duty cycle, the actuator is only required to exert a small force on
the target, to keep it levitated out of the beam.
At the appropriate time, a large accelerating force is required over a short period 
to accelerate the target into and out of the beam and then bring it to rest
at its levitated holding position.
For this short period, high currents can be employed.

The motor must be of a permanent magnet, brush-less design, as the high acceleration 
and large travel of the motor rule out the placing of coils on the moving parts.
The integration of permanent magnets into the moving assembly removes the need for 
electrical contacts between the stator and the moving parts, simplifying the 
interface between the motor and the ISIS vacuum.
The magnets on the moving components interact with the field produced by a set of 
stationary coils in the stator body.
These coils are outside the ISIS vacuum, directly wired to the driving electronics.
Positioning outside the vacuum also allows the use of a water cooling circuit to
remove the energy deposited by Joule heating of the coils.

The initial design of the linear motor was based on studies performed by an 
electrical engineer specialising in motor design, and outlined in \cite{Schofield}.
The important constraints were to maximise the accelerating force while minimising 
the mass of the moving components.
The mass of the magnetic materials thus form a significant fraction of the total 
mass.
Different magnet and coil topologies were investigated, and the resulting design
is documented in the following sections.

%
\subsection{Electromagnetic design}
\label{Sect:LinearMotor:EMDesign}
%
Analysis of ISIS beam properties indicated that a peak acceleration of 780\,m\,s$^{-2}$
would give sufficient headroom for the target to achieve an appropriate interception with the beam
given various beam conditions and a deep target actuation.
The low mass of material required for interaction with the beam implies that the mass of the moving
part of the motor (or ``shuttle'') must be dominated by that of the permanent magnet assembly and 
any mechanical linkages.
A design was therefore required which maximised the electromagnetic force while minimising
the mass, with the goal of a specific force equal to $\sim$780\,N\,kg$^{-1}$ for reasonable 
assumptions of motor geometry and location.
To achieve the highest magnetic loading, sintered neodymium-iron-boron magnets were chosen
for the shuttle, as these have the greatest field strength and best
strength-to-mass ratio.
A bank of appropriately energised coils interact with the field of the magnets to drive 
the shuttle.
Soft magnetic core material was considered for the stator, but this was not found to lead
to any advantage, due to the small size of the motor
and the fact that magnetic material would be saturated \cite{Schofield}. 
This was exacerbated by the significant ``air-gap'' between the permanent magnets
(inside the vacuum chamber) and coils (outside), actually filled by vacuum and non-magnetic 
vacuum tube.

Two magnetic topologies were considered for the shuttle, and compared using 2-D axisymmetric 
modelling.
Multi-pole radially magnetised discs attached to a central soft magnetic core  were found to 
provide a more efficient device than axially magnetised discs separated by pole pieces.
The radial design is also less prone to demagnetisation \cite{Schofield}.
The exact geometry was then improved by iterative finite element studies.
Test results from a prototype motor were used to validate the design.

%
\subsection{Stator}
\label{Sect:LinearMotor:StatorCoils}
%
%
The stator, which is cylindrical in shape, contains a set of flat coils mounted 
around a thin-walled steel tube.  
After winding, each coil is impregnated with insulating varnish to form a 
stable compact unit. 
During assembly six 25\,${\mu}$m copper shims are sandwiched between  
each pair of coils to facilitate heat conduction out of the coil stack.
The addition of the copper shims gives a coil pitch of 3\,mm. 
Connecting leads from the coils are led radially outwards. 
Three thermocouples are inserted between three pairs of coils to enable the 
temperature of the coil stack to be monitored. 
A coiled copper tube soldered onto a solid copper jacket is placed around the 
coils and is in contact with the copper shims. 
This carries the cooling water, the temperature of which is monitored 
at either end with two more thermocouples.  
The entire assembly is inserted into an aluminium outer cylinder, the stator 
body, with the insulated copper wires and the cooling pipes emerging through 
a slit in the side. 
The individual coils are wired up at terminal blocks placed external to the 
stator body.

\subsubsection{Coils}

The stator contains twenty-four identical coils that are stacked
vertically and numbered one to twenty four starting from the top of
the stator. 
The stator coils are responsible for interacting with
the permanent magnets on the shaft both to levitate the target shaft when the 
target is being held out of the ISIS beam and to produce the
accelerating force when the target needs to be inserted into the beam. 
(A motorised jacking platform is used to raise the target from the beam when 
not in use, as described in section \ref{Sect:Mechanical:Integration}.)

\begin{figure}[!htb]
\begin{center}
\includegraphics[height=8cm]{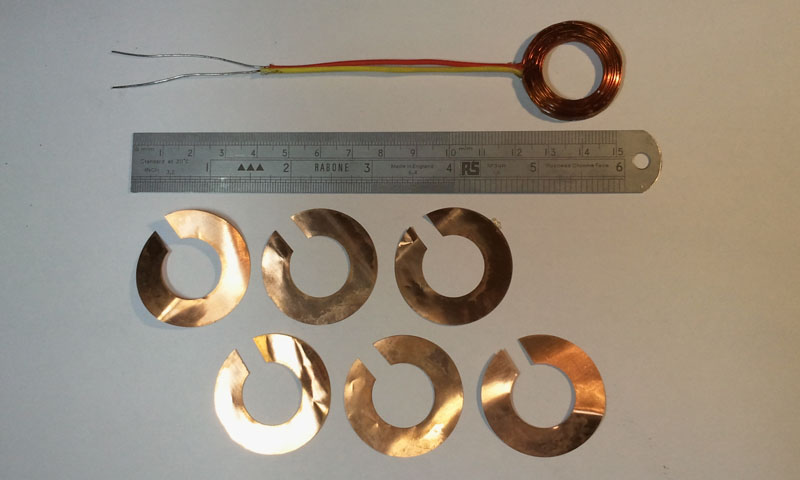}
\caption{A complete stator coil and some copper heat-sinking shims.} 
\label{PhotoCoil}
\end{center}
\end{figure}

Each coil is composed of thirty six turns of 0.56\,mm polyester-imide
enamelled copper wire, over-coated with a polyamide-imide resin. 
This yields a high temperature winding wire that is rated to 200\degC
operation\cite{WireDS}. 
These coils are wound on an 18.1\,mm diameter former, each coil having 
a depth of 2.85$\pm$0.1\,mm. 
Coils outside this tolerance are rejected due to the limited space between 
coils and the required pitch of 3\,mm. 
The clearance of $\sim$0.15\,mm between the coils 
is used to insert thin copper shims which act as heat sinks.
After winding, the outer diameter of the coils is 30\,mm. 
Each coil is double dipped into a varnish that seals the windings and 
provides additional electrical insulation\cite{Varnish}.
Each coil is tested to 1\,kV before being built into a stator. 
A photograph of a finished coil and some of the inter-coil copper shims is
shown in figure~\ref{PhotoCoil}.
The copper shims provide a thermal path between the coils and the cooling jacket.
The inner diameter of the shims is 19\,mm and the outer diameter 36\,mm.
The shims are split to reduce eddy-current losses.

\subsubsection{Stator Bore}

A thin-walled non-magnetic stainless steel tube passing through the centre of 
the coils forms the stator bore.
It provides isolation between the stator body and the ISIS vacuum and ensures
the mechanical alignment of the coil stack.
The nominal wall thickness of 0.5\,mm is reduced to 0.3\,mm where it passes through 
the coils.  
The reduction in magnetic field strength within the bore caused by the 
stainless steel tube was estimated to be 1\%.
The stainless steel tube is insulated from the coil stack using three layers of
self-adhesive kapton tape.

\subsubsection{Cooling Jacket}

The cooling of the stator is extremely important as the rate of
heat transfer from the coils to the cooling water ultimately limits
the maximum rate at which the target can be actuated.
Typically, when the stator is levitating the shaft out of the beam,
the power consumption is {$\sim$}30\,W.
Every time the target is actuated an additional heat load of 400\,J 
is deposited in the stator coils. 
The coils are small and therefore the heat capacity of the coil stack is 
correspondingly low. 
Without any heat-sink the coils would rise in temperature by $\sim$ 5\degC 
with every actuation. 
Therefore, if this heat is not removed quickly, repeated actuation of the 
stator will rapidly result in the temperature of the coils rising above their 
maximum rated working temperature of 200\degC.

Unfortunately, the permanent magnets that are attached to the shaft will not 
operate up to this temperature without there being a serious risk of 
demagnetisation. 
The exact maximum safe operating temperature is hard to determine, as
the Curie temperature is field dependent. 
There is also some evidence that the risk of demagnetisation at elevated 
temperatures is accentuated when running permanent magnets in a radioactive 
environment\cite{Howlett:2007}.
Running the stator for extended periods has demonstrated that coil temperatures 
of 80--90\degC do not lead to demagnetisation.

\begin{figure}[!htb]
\begin{center}
\includegraphics[height=8cm]{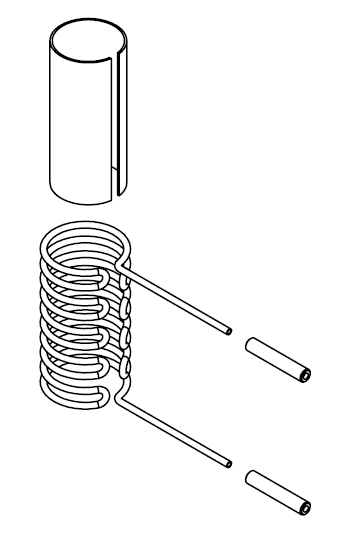}
\caption{The copper cooling jacket is over-fitted with the water cooling pipes.
The cooling pipes are soldered onto the jacket.} 
\label{Fig:JackDrg}
\end{center}
\end{figure}

A cooling circuit is required to remove heat from the coil stack.
This consists of an external, water-cooled, split cylindrical copper jacket.
The jacket has a thin-bore, copper cooling tube soldered onto its outer
surface through which a flow of water can be maintained.
This is illustrated in figure \ref{Fig:JackDrg}.

\begin{figure}[!htb]
\begin{center}
\includegraphics[height=8cm]{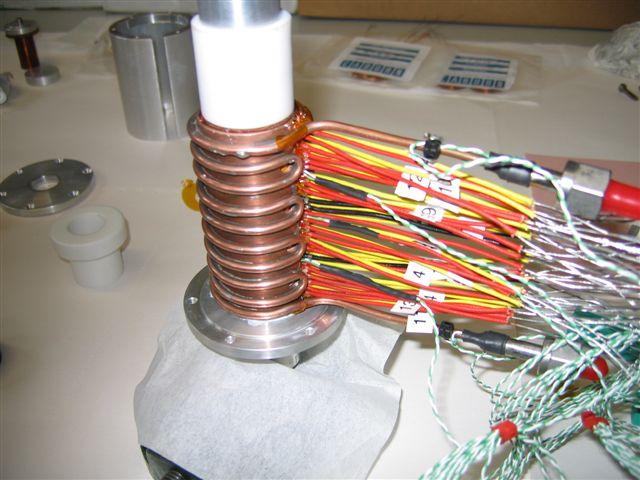}
\caption{The cooling jacket, placed over the coil stack.}
\label{Fig:JackPhoto}
\end{center}
\end{figure}

The inner diameter of the cooling jacket is slightly smaller than
the outer diameter of the copper shims.  
When the jacket is slid over the coil stack this has the effect of bending 
the copper shims over, thus ensuring a good thermal contact between them and the
jacket. 
A photograph of the jacket
placed over the coil stack is shown in figure \ref{Fig:JackPhoto}.
The cooling pipe has a narrow bore and so the water flow rate is quite low. 
Typically, at $\sim$4\,bar, a flow rate of $\sim$1\,litre\,min${}^{-1}$ is 
achieved. 
This flow rate has proved to be sufficient to remove the heat from 
the coil stack during normal operation. 
A nominal stator operating temperature of 80\degC has been maintained,
well within the working temperature range of all the components.

\subsubsection{Stator Assembly}

\begin{figure}[!htb]
\begin{center}
\includegraphics[height=8cm]{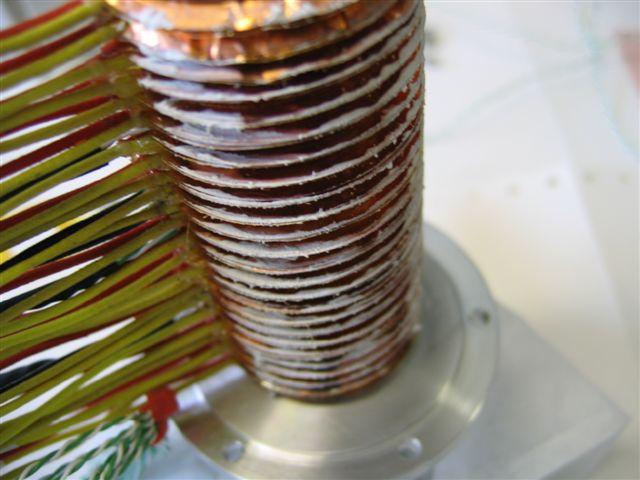}
\caption{All 24 coils assembled on a former. 
The copper shims can be clearly seen protruding from the coil stack where 
they will later make contact with the cooling jacket.} 
\label{StatAss2}
\end{center}
\end{figure}

The coil stack is assembled over a former that has the same
diameter as the insulated bore tube. 
The stack starts with a spacer followed by four copper shims. 
A coil, lightly coated on both sides with a thermal paste to aid with 
heat-sinking, is added. 
Six copper shims follow. 
The second coil is then added to the coil stack and the process is repeated 
for all twenty-four coils. 
As each coil has a nominal thickness of $2.85\pm0.1$\,mm the cumulative error 
is tracked and minimised during the coil stack construction by
selection of coils of appropriate thickness. 
An assembled coil stack is shown in figure \ref{StatAss2}.

The split in the copper shims aids in keeping the coils parallel
during construction. 
This is achieved by placing the split in the shim where the wires to the coil 
exit.
(There is sometimes a small bump at this point on the coil due to
the wire exiting from the centre of the coil back over the top of
the other windings). 
The gap created by the split in the shims also allows thermocouples to be
inserted into the coil stack.

At this stage a cooling jacket is slid over the coil stack. 
The stator body is then completed by adding a split outer jacket and two 
end-caps. 
The split in the outer jacket allows the wires and the cooling pipes to 
protrude for external connection. 
The end caps provide light compression on the coil stack, keeping it in place. 
The former on which the coil stack was formed is now removed and the 
bore tube inserted through the bore of the stator.
A final electrical insulation check is then performed, to ensure that all coils
remain isolated from metal parts including the bore tube and cooling jacket.
The installation of the stator body into the core of the target drive assembly
is described in section \ref{Sect:Mechanical}.

%
\subsection{Permanent magnets}
\label{Sect:LinearMotor:PermMag}
\begin{figure}[!htb]
\begin{center}
\includegraphics[height=8cm]{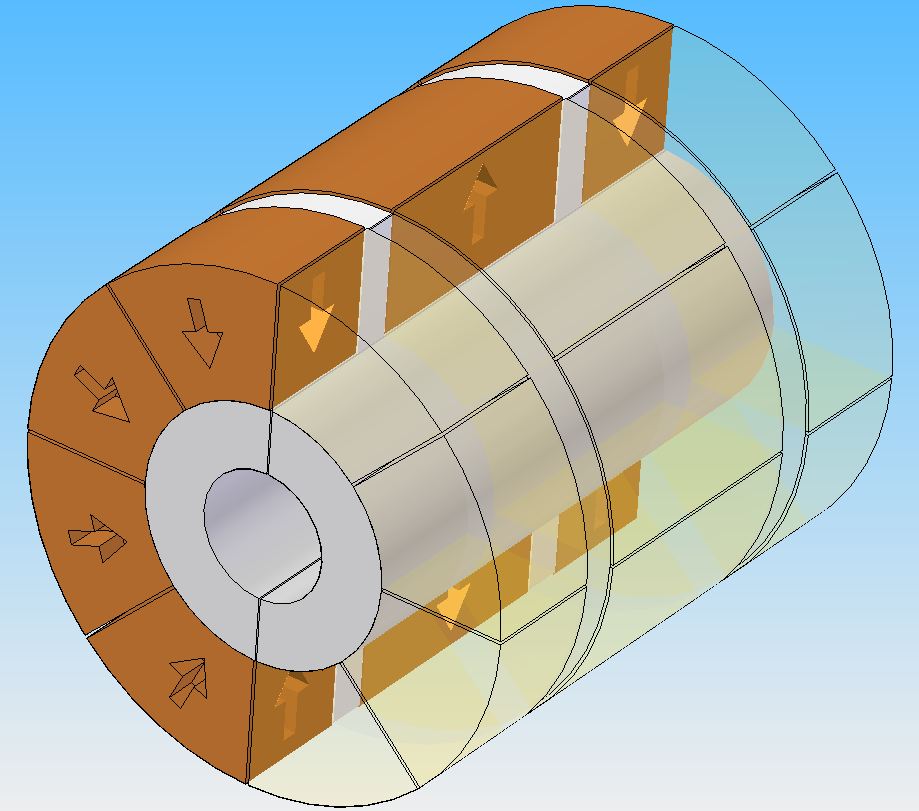}
\end{center}
\caption{Drawing showing the structure of the magnet
assembly}\label{fig:TDMagnet}
\end{figure}

%
%
The permanent magnet assembly interacts with the field of the stator coils
to produce the force on a central shaft which accelerates the target into
and out of the proton beam.
To achieve the maximum magnetic field, the assembly is constructed from 
sintered neodymium-iron-boron (NdFeB) magnets.
Twenty-four segments are arranged in 3 rings, with 8 magnets per ring, 
as shown in figure \ref{fig:TDMagnet}.
They are glued to a mild steel core to produce a cylinder that is 18\,mm
long with an outer diameter of 15\,mm and an internal diameter of 4\,mm.
Thin ceramic washers separate the three rings.
The central ring is 7.8\,mm long, twice the length of the two outer rings.
The whole assembly has a mass of about 25\,g.

The individual sectors are manufactured by wire erosion from 
un-magnetised NdFeB material.
The sectors are then magnetised radially and assembled in a jig before
finally being glued in position using a strong two-component epoxy.
Once the glue is cured, the magnet unit is lightly machined to the precise 
outer radius required.
The middle ring is magnetised so that the outer surface is a North pole, 
while the outer rings have the opposite polarity.

A FLUKA \cite{FLUKA} simulation has been performed to ascertain the radiation 
levels expected around the target.
Assuming 24\,hour operation at 1\,Hz for 180 days in a nominal year,
the expected radiation dose to the magnets is estimated to
be $\sim 1 \times 10^4$\,Gy.
To date, no evidence of degradation due to radiation has been observed.

%
\section{Mechanical design and construction}
\label{Sect:Mechanical}
%
%
%

The MICE target unit contains mechanical components and sub-assemblies that
are designed to: 
\begin{itemize}
  \item Combine accurately all mechanical,
        electronic, electrical-power and optical-readout functions
        into a single unit;
  \item Provide a vacuum tight volume connected hermetically to the ISIS beam-line;
  \item Enable target operation by:
   \begin{itemize}
    \item Providing controlled drive of the target's shaft into and out
          of the beam;
    \item Enabling the velocity and location of the shaft to be determined;
    \item Constraining the shaft along its path of travel thus preventing
          significant off-axis movement;
    \item Providing stiffness in the shaft that resists significant deformation
          and vibration during operation.
   \end{itemize}
  \item Eliminate the possibility of a failure that prevents continued
        operation of ISIS beam, including:
    \begin {itemize}
     \item Breaking of the shaft such that part falls into the circulating beam;
     \item Failure of welds or seals causing a leak into the ISIS vacuum chamber; and
     \item Contamination of the ISIS vacuum chamber, particularly with metallic or other dust that
           could contaminate the RF cavities.
    \end{itemize}
   \item Deliver an operational lifetime that is sufficient to allow a rolling programme of 
     maintenance at appropriately infrequent intervals.
\end{itemize}

\begin{figure}
  \begin{center}
\includegraphics[height=8cm]{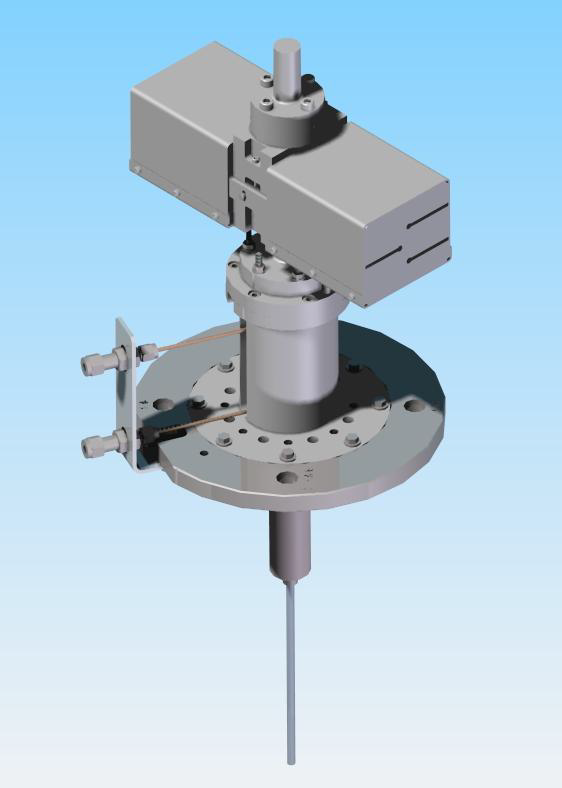}
\includegraphics[height=8cm]{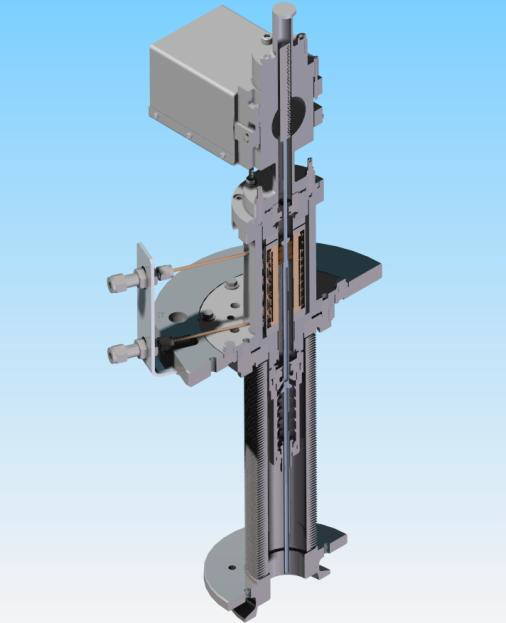}
  \end{center}
  \caption{
MICE target assembly, (left) with bellows removed, (right) a cutaway view
(Descriptions and information on the components in the assemblies shown here 
follows.)
  }
  \label{Fig:Mechanical:T1}
\end{figure}

%
\subsection{Target shaft}
\label{Sect:Mechanical:Shaft}
%
%
The shaft comprises of two sections of titanium (Ti) alloy grade 5
(6\% Al, 4\% V).
The sections are a solid upper section and a lower tubular section with
various functional features on each.
The two halves are joined with a shrink-fitted plug-and-socket arrangement then 
electron-beam welded together. 
The shaft is coated with a hard diamond-like carbon (DLC) coating that acts as
a bearing surface.
The shaft is finally fitted with a permanent magnet, a slotted graticule for
position read-out and the associated fixings and fasteners.

\begin{figure}
 \begin{center}
\includegraphics[height=4cm]{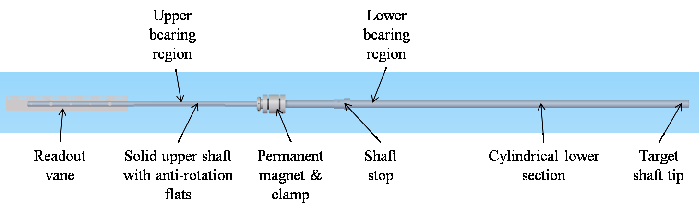}
 \end{center}
  \caption{
    MICE target shaft assembly (shown horizontally oriented, operates
    vertically with tip down).
  }
  \label{Fig:Mech:ShaftAss}
\end{figure}
 
The shaft is 528\,mm long and, when fully assembled with the magnet, slotted
graticule and fixings, has a weight of $\sim 51$\,g.
A radially segmented, permanent-magnet assembly
is bonded to the shaft and held with a mechanical clamp of minimal mass
(1.5\,g).
The stator accelerates and then decelerates the shaft in both the down-stroke 
and the up-stroke with a maximum travel in each direction of around 48\,mm.
The shaft achieves this 96\,mm of reciprocating movement in about 30\,ms
before dwelling and then dipping again; with the dwell the shaft dips at a
frequency of just under 1\,Hz. 
The tip of the shaft has a cylindrical cross section of 
$\sim$11.5\,mm$^2$ (5.95\,mm outside diameter and 4.55\,mm inside diameter);
as it is an integral part of the tubular lower shaft it is grade 5 titanium alloy.
It is the tip of the hollow cylinder that momentarily grazes
the halo of the ISIS beam to produce pions.

The production quality of the target shaft is the most critical aspect of all
the manufacturing process.
The shaft is manufactured within very tight dimensional and geometric form 
tolerances. 
The tight dimensional tolerances ensure the close fit of the shaft with the 
bearings; this tight fit accurately constrains the shaft's vertical motion. 
The shaft requires a high
degree of straightness and good roundness. 
Tight tolerances of form (cylindricity and run-out) are applied to the 
manufacture of the shaft; accurate form minimises deformation and vibration 
of the shaft during the rapid acceleration and deceleration.

\subsubsection{Design}
Though form is accurately controlled, it is not perfect.
Add to this a non-perfect constraint in the bearings and a stiffness limited by
mass, geometry and material properties, then
the rapid acceleration and deceleration of the shaft will cause it to deform 
and vibrate during operation. 
The stiffness of the shaft is very important. 
Increasing the stiffness of the shaft, without increasing mass, pushes up 
the frequencies of resonant modes. 
Increasing the frequencies, especially those of the first few resonant modes, 
reduces the number of resonant modes the shaft passes through as it is 
accelerated and decelerated. 
Excitation of vibrations in the shaft may arise from a number of sources, including:
\begin{itemize}
   \item Large changes in velocity take place at $\sim 15$\,ms intervals, 
    i.e. twice over the 30\,ms cycle.
    If the natural frequency of the shaft is close to 33\,Hz or 67\,Hz 
    some distortion will be induced.
   \item The switching of power from one set of coils to the next 
     may introduce an excitation force. 
     The longitudinal driving force (up to 50\,N) may be accompanied by off-axis
     torque or lateral forces as the sequential coil-switching takes place
     during dipping. 
     The frequency and severity of the excitation would be directly linked to 
     the timing and magnitude of these forces. 
     The timing of the switching is variable as the shaft is accelerated and 
     decelerated in both directions, so the frequency of excitation is 
     variable too.
     This may cause the shaft to pass through several excitation frequencies 
     that correspond to the shaft's resonant frequencies. 
     The magnitude of the excitation depends on the force seen by the shaft
     when switching between coils; this is related to the
     quality of the coils and their axial location relative
     to the permanent magnet. 
     With a 0.5\,mm offset in coaxiality between the permanent magnet and the 
     coil's magnetic centre the lateral forces have been calculated to be up to 2.5\,N 
     and the torque up to 32.5\,mN\,m. 
   \item The unit or surrounding support frames may be excited by the 
     dipping shaft. 
     There is potential for this vibration to transfer back from the support 
     frame to the shaft through the stator and the bearings. 
     There is significant vibration felt on the test frame after each dip.
   \end{itemize}

A tubular, `O' section shaft was shown to have sufficient stiffness.
A finite-element (FE) modal analysis showed the 1st and 2nd resonant modes at 50\,Hz, but 
the 3rd and 4th at 68\,Hz.

\begin{figure}
 \begin{center}
\includegraphics[height=5cm]{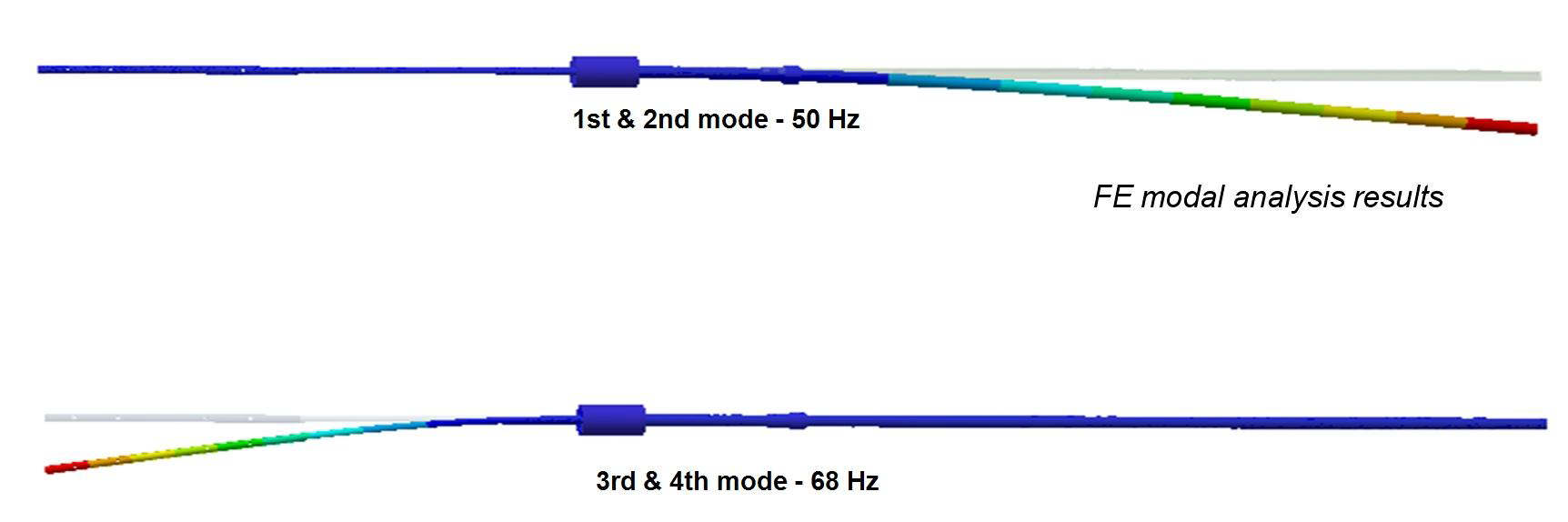}
 \end{center}
  \caption{
    Output plot of FE modal analysis of shortened shaft with `O' shaped lower
    section.
  }
  \label{Fig:Mech:FEmodal}
\end{figure}

The tubular lower section is created by rough-machining a bar to leave a 
slightly oversized external profile, including an integrated mechanical end-stop. 
This is then heat-treated to remove residual stresses. 
The bore is then wire eroded to 4.55\,mm ID over the length of the lower shaft 
(about 320\,mm); if internal stresses were present the shaft might deform along 
its length as this material is removed.  
After wire erosion the external profile, including the stop, is finished to final 
size referenced to the bore to maintain concentricity and thus material 
balance.  
Finally the socket for the shrink-fit joint is added.

The upper shaft stock material is stress-relief heat-treated before manufacture.
It is then ground; this includes the flats which are very finely ground to 
ensure internal stresses are minimised.  
A larger diameter is left at one end onto which the socket for the shrink-fit
joint is machined.  
Also added are the undercut for the magnet clamp and the fixing holes for the 
laser readout graticule.  
The very fine slot for the graticule is then cut by wire erosion.
The wire erosion of this slot causes the two adjacent sections to open in a
`Y' shape.
(This is only seen on the grade 5 titanium shafts and not on the previous versions
made from commercially pure Ti grade 2; in fact the slot collapsed slightly when
using Ti grade 2.)
It appears the wire cutting causes a slight expansion in the surface
material on the inside of the slot that forces these sections to fan out.
This has been resolved by supporting the shaft with the slot closed and 
stress-relieving back to a correct form.

\begin{figure}
 \begin{center}
\includegraphics[height=8cm]{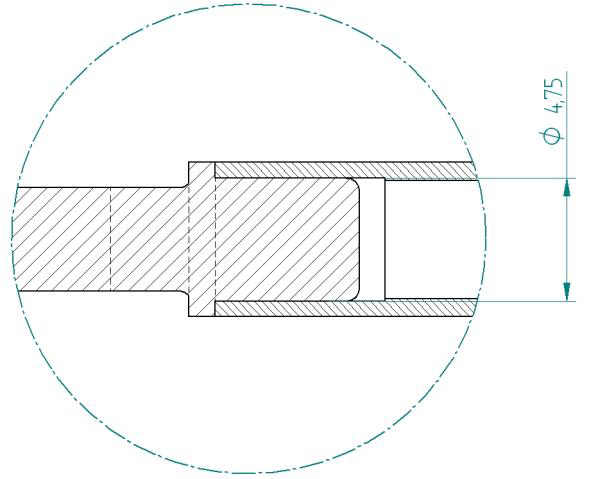}
 \end{center}
  \caption{
    Plug and socket detail, upper shaft on left, lower shaft on right
  }
  \label{Fig:Mech:PlugSocket}
\end{figure}

To alleviate the risk of undertaking so many operations on a long single-piece 
shaft it is produced in two sections.
The additional benefit of this is that they can be produced concurrently with 
benefits to the schedule.  
The lower shaft has an accurately machined socket and the upper 
shaft has a mating plug. 
The plug and socket are manufactured with a nominal 0.01\,mm interference press 
fit. 
If these were pressed together at ambient temperature the material would cold-weld.
Cold-welding is unlikely to be uniform and so the joint would skew causing a 
significant form error between the upper and lower shaft sections; it may even 
seize before it is fully pressed together.  
To achieve a clearance as they are assembled (i.e. to prevent cold welding or 
galling) the socket is expanded by heating to 250\degC and the plug is shrunk 
by cooling to $-190$\degC in nitrogen.
A precision jig is then used 
to ensure correct alignment as the pieces are pushed together. 
The interference-fit is made as the joint comes back to room temperature, 
contracting the socket and expanding the plug.

The two halves are further secured by welding at 8 points along the seam 
between the plug and socket using electron-beam (EB) welding. 
A full circumferential weld was originally tried, however under-run or over-run
at the end of the weld always made the two halves skew too much.  
Tensile testing was used to determine the strength of the shrink-fit and
point-welded joint.  
Tensile testing broke the spot-welds at 4\,kN which coincided with an audible 
crack as the welds broke. 
After the welds broke the interference-fit of the plug and socket prevented 
the shaft breaking into two halves; instead the tensile load began to climb 
in finite steps against displacement to 6\,kN before the test was stopped.
This was due to a cold-welding effect; this is beneficial as it significantly 
reduces the chances of the shaft breaking into two should high impact load 
occur during a fault.

\begin{figure}
  \begin{center}
\includegraphics[height=4cm]{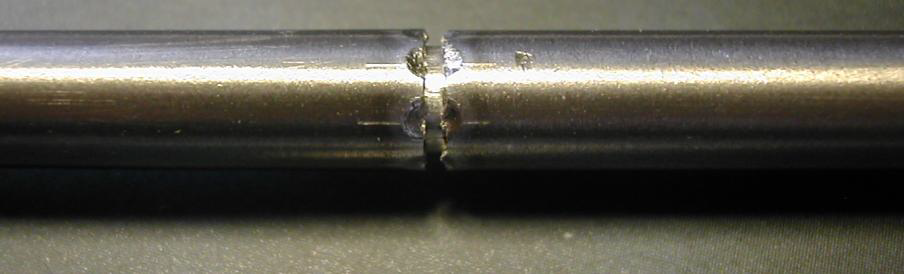}
  \end{center}
  \caption{
    Failed weld and displacement of shaft halves after tensile testing
  }
  \label{Fig:Mech:FailedWeld}
\end{figure}

\subsubsection{Material}
Titanium is a good target material with respect to particle production, it is 
widely available, it can be easily worked and can be joined by welding.
The shaft needed to be low mass to enable rapid and efficient acceleration by 
the stator. 
The shaft also needed to be stiff to minimise elastic deformation and 
vibration during its operation; titanium and its alloys have a good
stiffness-to-mass ratio.

\begin{figure}
  \begin{center}
\includegraphics[height=8cm]{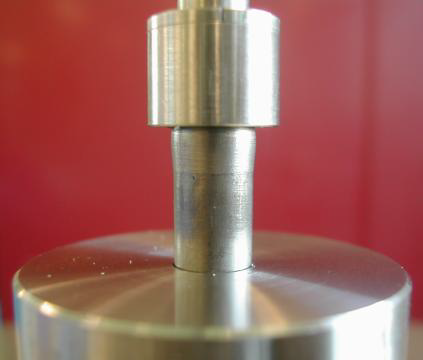}
  \end{center}
  \caption{
    A bulge can be seen in the shaft under the stop. 
    This occurred during impact testing.
  }
  \label{Fig:Mech:Bulge}
\end{figure}

The initial redesign of the lower shaft led to the use of a tube with a 
mechanical stop welded on; tube in unalloyed grade 2 is widely available
and was initially chosen for the shaft material.
Impact testing was undertaken on this welded-stop design; this closely 
represented a shaft's mechanical end stop impacting the lower bearing at a 
maximum speed of 9.3\,m\,s$^{-1}$ during a fault condition, including a
safety factor of 1.5.
A mass was dropped from a set height onto the stop on the lower shaft.
This stop failed at around 4 impacts.
In addition a bulge started to form in the wall of the tube due to the
impact, see figure \ref{Fig:Mech:Bulge}.
Though the EB welding is a semi-automated process there is potential for 
variability in the strength of the weld, so 
the welded-on stop was not an entirely risk-free choice.  
In addition the welding of the stop caused some distortion to the tube that 
had to be corrected by mechanical manipulation. 
It was decided to pursue a more costly manufacturing route to produce a 
stop integrated into the lower section from a single bar, a much stronger 
design.  
This overcame the restriction of obtaining only grade 2 tube and allowed 
consideration of titanium alloys.

Grade 5 (Ti-6AL-4V) titanium alloy was eventually chosen.
There is a slight improvement in stiffness with the grade 5 titanium alloy 
over the unalloyed grade 2 titanium, 114\,GPa and 103\,GPa respectively.
The additional strength of the grade 5, 895\,MPa minimum over the 395\,MPa minimum of 
grade 2, is of no significant benefit as the loads on the shaft are minimal 
during normal operation and the integrated-stop design is much stronger.  
The biggest benefit of grade 5 over grade 2 is in the hardness -- Rockwell C 
36 and 21 respectively in an annealed state.  
The harder alloyed grade 5 is easier to grind, allowing better finishes 
and tighter tolerances to be achieved.
This is particularly useful for the upper shaft where the diameters and 
flats are ground.

\subsubsection{Manufacture}
Grinding is used to produce the main geometry of the upper shaft. 
The depth of cut in the grinding process is reduced towards the 
final cut to minimise the amount of deformation induced into the surface of 
the component; this in turn minimises internal stresses in the final 
shaft. 
Internal stresses may relax via strain over time leading to warping of 
the shaft. 
Grinding also gives the shaft a fine surface finish so that polishing 
removes only a minimal amount of material in achieving the final surface 
roughness of 0.05\,Ra.

Stress-relief annealing is undertaken on raw materials before processing as
well as at certain stages throughout the processing.
This removes internal stresses induced by manufacturing processes.
Unrestrained heat treatment to around 670\degC with a dwell period of $40-50$
minutes followed by a slow cool relaxes the stresses through strain, which deforms the 
material slightly.
This deformation is removed during subsequent processing to achieve the final 
component's size with minimum internal stresses.
The temperature and heat treatment cycle information was taken from \cite{DIN65084}.

\begin{figure}
  \begin{center}
\includegraphics[height=8cm]{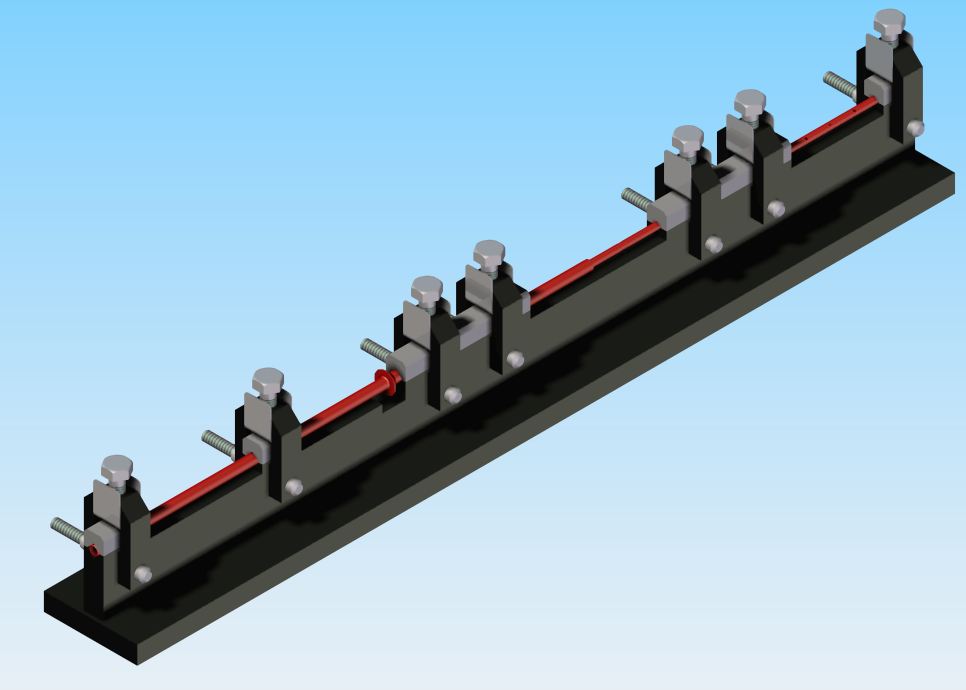}
  \end{center}
  \caption{
    Shaft in straightening jig
  }
  \label{Fig:Mech:Jig}
\end{figure}

A stress-relief anneal is also applied to the assembled shaft prior to the 
final DLC coating process to straighten it.  
The shaft is held firmly and accurately in a jig, see 
figure \ref{Fig:Mech:Jig}.  
The jig is made from mild steel which is itself stress relieved (at 700\degC) 
before an accurate `V' is finely ground into it that supports the shaft.  
Mild steel has a similar coefficient of expansion to titanium alloy, 
11\,ppm/\degC and 9\,ppm/\degC respectively; this means during expansion and 
contraction upon heating and cooling there will be minimal force induced by 
the jig on the shaft.  
Mild steel is also a dissimilar material to the titanium alloy so there is less
chance of the jig and the shaft bonding at the elevated temperatures during 
heat treatment. 
The latest shafts have all undergone this process and have been straightened 
from \textgreater 0.5\,mm run-out post manufacture to around 0.1\,mm run-out post treatment.

Conventional machining, such as milling and turning, is used on many features
including the profile of the lower shaft, the fixing holes for the 
graticule and the plug-and-socket features used to join the upper and lower 
shaft sections. 

Wire-erosion is used to create a  90\,mm by 0.22\,mm slot in the  
upper shaft to support the graticule.
The 4.55\,mm inner diameter, 320\,mm long bore through the lower shaft is
also made by wire-erosion.

The sub-components of the shaft, and the fully-assembled shaft, are inspected at 
many stages during manufacture to determine if they meet the specification on 
the technical drawings. 
Further information on the inspection of the shaft is given below. 

Diamond-like carbon (DLC) coating is applied to the bearing surfaces of the 
target shaft.  These surfaces are polished prior to DLC coating to 
ensure the final finish is smooth.  
The tip of the shaft that dips into the beam is not coated.  
The tip is left uncoated to prevent any thermal heating, from contact with the 
ISIS beam, causing failure of the coating through high temperature 
(\textgreater 400\degC) or through thermal strain differentials.

\begin{figure}
  \begin{center}
\includegraphics[height=8cm]{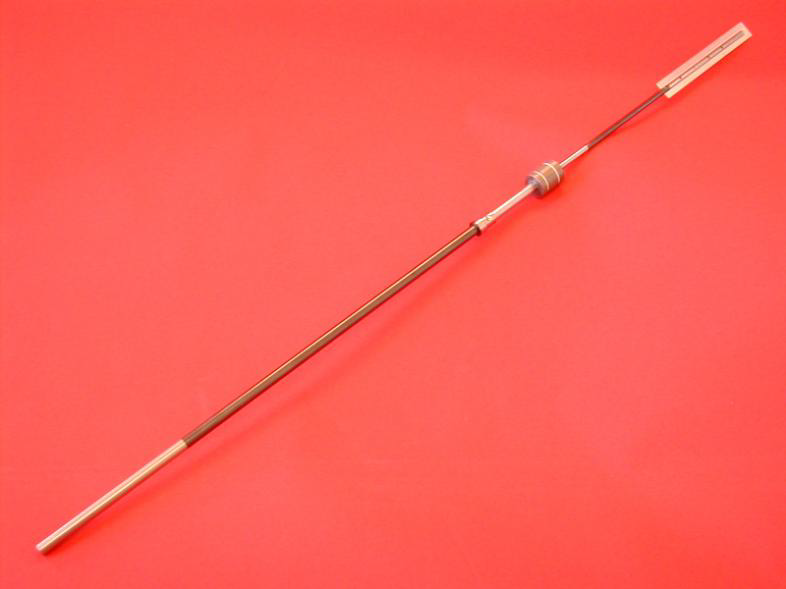}
  \end{center}
  \caption{
    DLC coated shaft, dark grey/black coating extending over and beyond the 
    bearing regions of shaft.
  }
  \label{Fig:Mech:DLC}
\end{figure}

\subsubsection{Shaft measurement}
The shaft cannot be made perfectly straight due to its 
long thin shape,
but as long as the form is controlled within tolerable limits operation 
will be acceptable. 
During manufacture care is taken to produce the shaft sections with the best 
possible straightness, minimising run-out between the bearing regions 
as well as finely balancing material about the axis of the shaft.  
Manufacturing includes the use of jigs to hold the shape during processing, 
heat-treatment stages to relieve stresses and a final constrained 
heat-treatment to achieve the desired shape.
Shafts are accepted with a run-out between bearing regions of less than 
0.12\,mm; through prototyping it was proved that shafts with such
run-out operated acceptably in the target unit.

The size of the shaft is very easy to measure using conventional micrometers.
The shape however is more difficult.
For the shaft a Taylor Hobson Talleyrond is used to measure the run-out 
between the lower bearing region and the other parts of the shaft.
The Talleyrond has a rotating table onto which annular parts are clamped; a 
probe on an extending arm then measures the exterior of the annular component 
as the table rotates.  
A jig was produced for shaft measurement that allows the Talleyrond 
to access relevant sections of the shaft for measurement.  
It uses an upright post to support a collet assembly; by lightly press-fitting 
the shaft into the collets, then the collets into a holder, the shaft is secured 
and aligned.  
The larger diameter lower bearing region is chosen to be the datum from which all
other surfaces are measured.  
The Talleyrond is programmed to locate this surface by measuring the 
circumference over a number of positions along its axis.  
It then aligns the lower bearing region with the axis of rotation,
by automatically adjusting the plane of the table.  
Next it measures the other sections of the shaft.  
Again the Talleyrond measures the circumference over a number of places along 
the axis of the region to produce a profile of slices. 
These 2-D slices can be stacked to produce an extrapolated 3-D model of
the shaft as well as being referenced to the lower bearing datum.
The probe has an extremely light touch (1.5\,mN), but this is enough to 
bend the shaft elastically when measuring further away from the collet supports.
A fine sprung-arm counter-balance is placed to oppose the probe arm to 
counteract any forces from the probe during measurements.

%
\subsection{Target bearings}
\label{Sect:Mechanical:Bearings}
%
%
\subsubsection{Description}
\begin{figure}
  \begin{center}
\includegraphics[height=8cm]{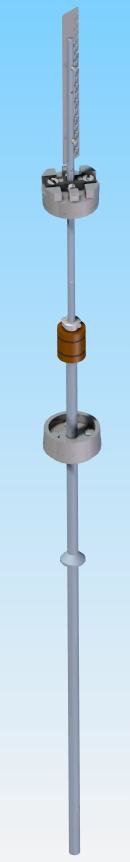}
\includegraphics[height=8cm]{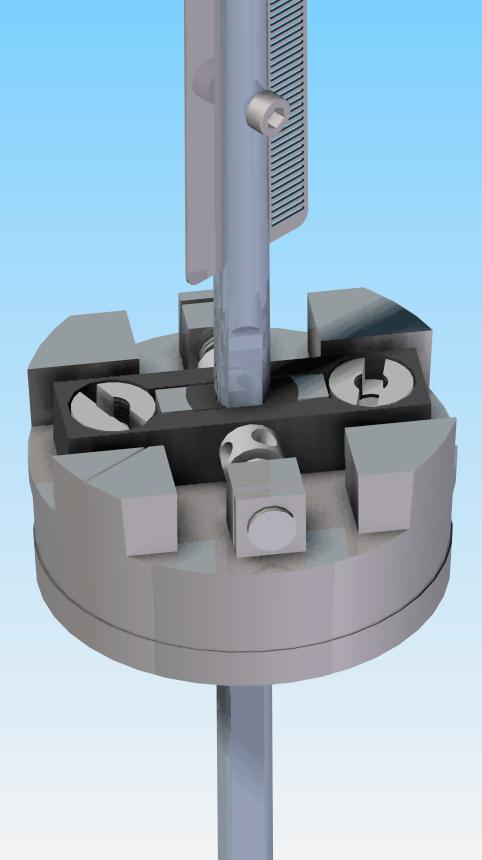}
\includegraphics[height=8cm]{04-Mechanical/Figures/JT12b.png}
  \end{center}
  \caption{
    (left) Shaft and bearings, (middle) Upper bearing with anti-rotation, 
    (right) Lower bearing
  }
  \label{Fig:Mech:ShaftBearings}
\end{figure}

Plain bearings support and constrain bearing 
regions on the shaft. 
There are two bearings, one to restrain the upper section of shaft and one to 
restrain the lower section. 
They are required to constrain the lateral motion of the shaft as it dips, as 
well as incorporating an anti-rotation feature to prevent significant rotation 
about the longitudinal axis of the shaft. 
The upper bearing has the anti-rotation component; it is finely adjustable via 
screws to ensure accurate location with respect to the anti-rotation flats 
on the upper shaft. 
The bearings have plastic (Vespel\circledR\, SCP5000) inserts and anti-rotation 
components (the dark items in figure \ref{Fig:Mech:ShaftBearings}). 
The plastic parts are assembled into stainless steel (Nitronic\circledR\, 60) 
carriers.
Nitronic\circledR\ was chosen as it minimises cold welding when the fine 
location taper on the outside of the bearing carriers are lightly pressed into 
the tapered bearing seat in the 1.4404~/~316L stainless steel body.

\subsubsection{Material}
For the bearings the material and coating combinations at the bearing 
interface with the shaft were very important for function and longevity.  
A number of material and coating combinations have been tried
for use on the target. 
From the early trials of different coatings, a diamond-like carbon (DLC) coating
on both the shaft and the bearing was found to be suitable. 
Despite these early successes, there subsequently appeared to be too 
much variability in the quality of this coating. 
This may have been due to the geometry of the parts being coated; the small 
bearing hole, even if produced in several segments, produced a ``hollow 
cathode effect'' in the RF coating chamber, thus variability in the thickness 
and adhesion of the DLC coating. 
This variability led to coatings on some shafts and bearings wearing away 
rapidly in later trials.

The alternative to DLC came from cryocooler technology as designed and built 
by STFC's Cryogenics and Magnetics Group. 
The cryocoolers have demonstrated in excess of 10$^{10}$ cycles in a dry helium 
atmosphere. 
The material combination used in these cryocoolers is titanium (both 
commercially pure and 6\% Al, 4\% V) against Vespel\circledR\ SP-3 grade. 
The SP-3 grade of Vespel\circledR\ is a polyimide that contains 
molybdenum-disulphide (MoS$_2$). 
MoS$_2$ is a dry lubricant which lowers the coefficient of friction. 
However MoS$_2$ was rejected on the grounds that long-lived isotopes might be
produced in the high radiation environment. 
DuPont$^{\rm{TM}}$ Technical were contacted to determine if an unfilled alternative 
may be suitable. 
From the following requirements it was determined that Vespel\circledR\ SCP5000
would meet the needs of the target when operating on the ISIS beam-line.

Bearing surfaces have a higher coefficient of friction in the ISIS vacuum of
10$^{-7}$\,mbar \cite{ISIS_Vac},
i.e.\ without the presence of moisture or lubricating films. 
A high coefficient of friction between bearing surfaces will increase the power
required to drive the shaft, it will cause accelerated wear of the bearing 
surfaces and it will raise the temperature at the bearing interface through 
frictional heating. 
SCP-5000 was recommended for use as it has a relatively low coefficient of 
friction, even in a vacuum environment, of 0.26 or better \cite{Vespel}.

Bearing materials are required to be tolerant to the effects of the nuclear
radiation generated in ISIS, i.e.\ their mechanical properties must not change 
significantly during the operational life of the unit.  
Vespel\circledR\ is a polyimide which will tolerate a significant total 
accumulated ionising radiation dose before a loss in mechanical properties.

\begin{figure}
  \begin{center}
\includegraphics[height=8cm]{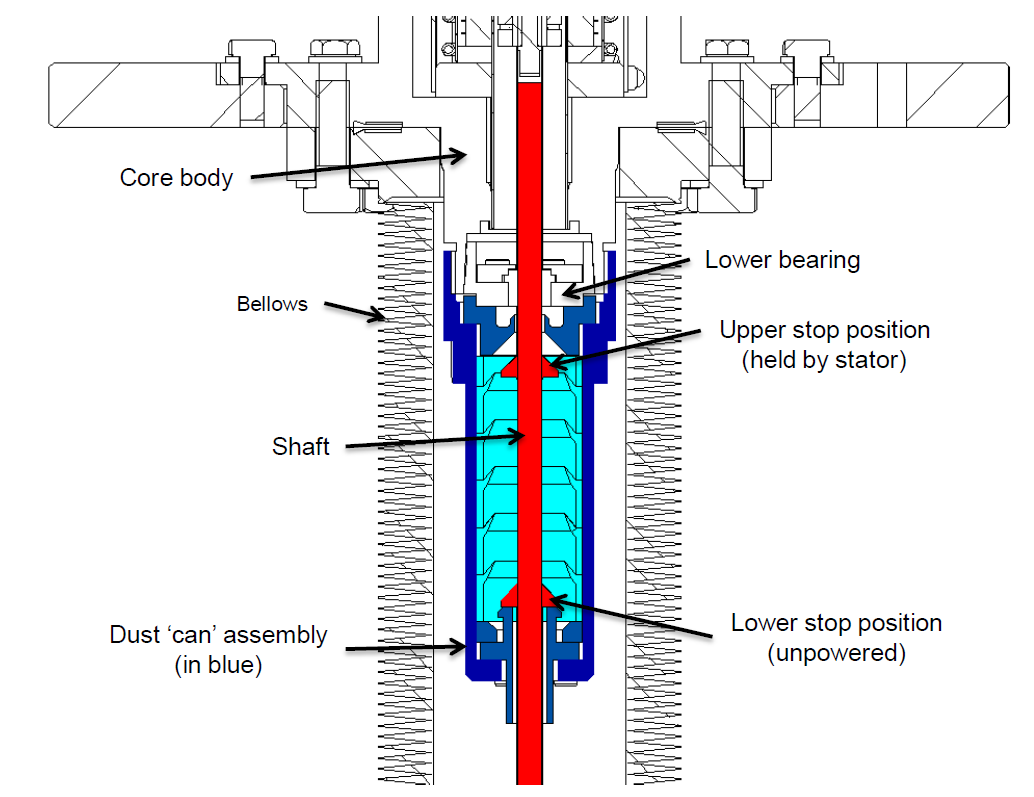}
  \end{center}
  \caption{
    Dust `can' assembly, that captures the dust below the lower bearing.
  }
  \label{Fig:Mech:DustCan}
\end{figure}

Particles created during wear should be kept to a minimum. 
If they are created they must be managed and contained within the unit. 
This is particularly important for particles that have become activated, for 
health and safety reasons.  
In addition particles should not be released from the unit 
as they may damage the RF cavities adjacent to the MICE target in ISIS.  
Dust is produced when using the Vespel\circledR\ bearings, however with a 
finely polished shaft this has been minimised.  
The target unit has been fitted with an extended housing with dust-traps to 
prevent the dust produced from the Vespel\circledR\ escaping outside of the 
unit, see figure \ref{Fig:Mech:DustCan}. 
The bearing surfaces need to withstand a significant operational life and be 
able to be manufactured with fine surface finishes and tight tolerances to 
minimise the rate of wear; Vespel\circledR\ is a relatively hard plastic so 
suits this.

The material must not significantly out-gas in the ISIS vacuum.  
Out-gassing tests were undertaken on vacuum-baked samples of Vespel\circledR\ 
SCP5000 (80\degC for 72 hours) which were shown to be acceptable for use on the ISIS 
beam-line provided the total volume was kept below 3 cm$^3$.
 
\subsubsection{Rotational constraint}
\begin{figure}
  \begin{center}
\includegraphics[height=8cm]{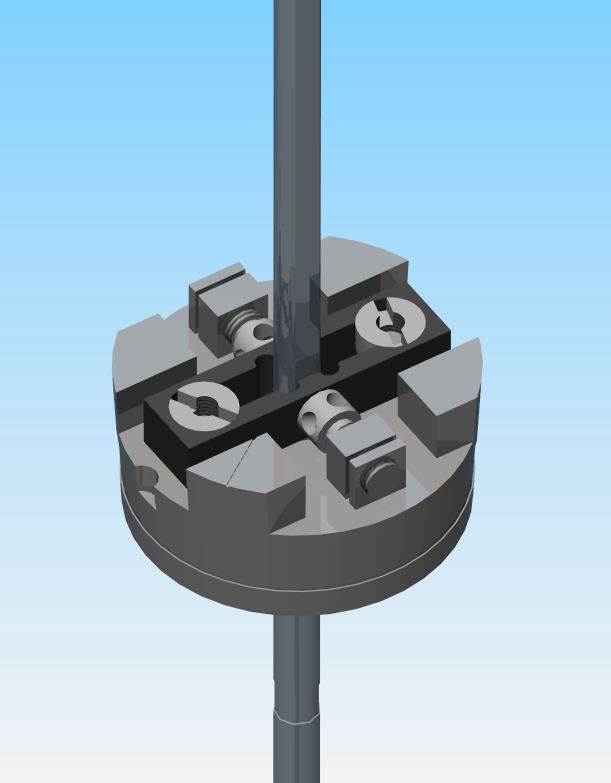}
  \end{center}
  \caption{
    Upper bearing with 
    Vespel\circledR\, bearing insert and anti-rotation component
  }
  \label{Fig:Mech:Bearings}
\end{figure}

The position of the shaft is determined using an optical vane (section
\ref{Sect:OpticalPosition:OpticalVane}) which must be held roughly 
perpendicular to laser beams.
The orientation of the shaft must thus be restricted to a range of about
$\pm 10^{\circ}$.
The range is limited to less than $\pm 3^{\circ}$ by providing
the upper section of the shaft with a pair of finely-polished parallel flats 
that run against a flat-faced Vespel\circledR\ anti-rotation component on the 
bearing, see figure \ref{Fig:Mech:Bearings}.  
The flat faces of the anti-rotation component are finely adjusted so that the 
flats on the upper shaft only contact them if the shaft rotates. 
In trials using an anti-rotation component with two completely flat faces, 
it was found that after running for some time the corners of the flats on the 
upper shaft would dig-in to the plastic and cause the shaft to lock temporarily; 
the semicircular reliefs shown in figure \ref{Fig:Mech:Bearings} were 
added to prevent this.

%
\subsection{Stator}
\label{Sect:Mechanical:Stator}
%
%
\subsubsection{Description}
The stator assembly surrounds the middle portion of the shaft assembly, 
including the permanent magnets which are driven by the coil stack. 
The bearings and stator assembly are fixed in the core structure (see below)
to coaxially align the shaft assembly and permanent magnet with 
the coils.

\begin{figure}
  \begin{center}
\includegraphics[height=8cm]{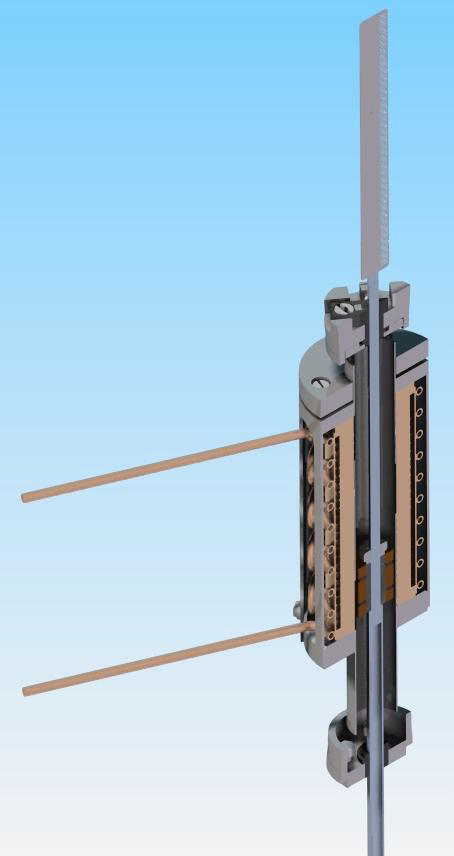}
\includegraphics[height=8cm]{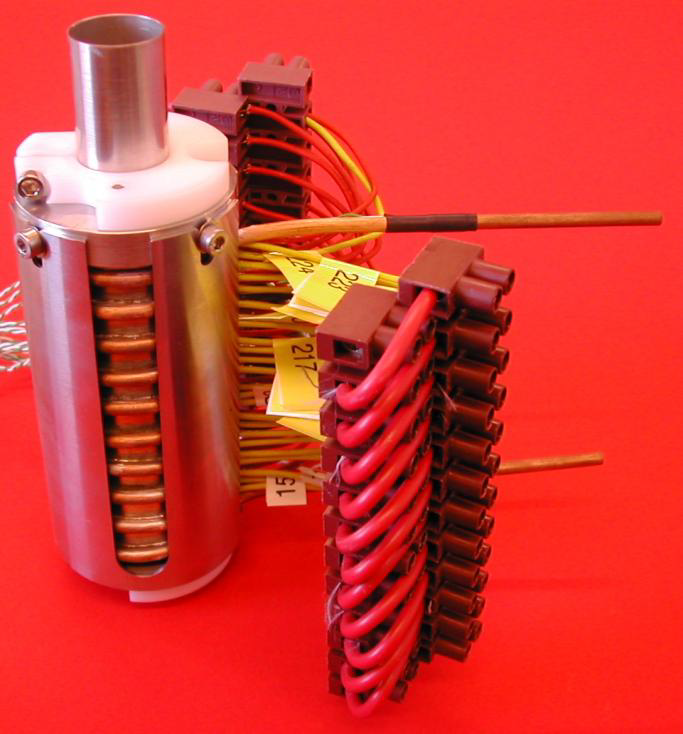}
  \end{center}
  \caption{
    (left) Stator assembled with shaft and disc spring plate on top, wiring 
    not shown; (right) Tested stator unit ready for assembly (shown with 
    temporary plastic clamps).
  }
  \label{Fig:Mech:Stator}
\end{figure}

The stator, described in section \ref{Sect:LinearMotor:StatorCoils}, is a 
self-contained sub-assembly; it is assembled around, but not 
fixed to, the central vacuum tube. 
As it is self-contained it can be tested independently.
After assembly and test the stator unit is incorporated into the target unit 
by welding the ends of the vacuum tube into the two main core body components. 
The tube and welds provide an uninterrupted vacuum volume though the middle of 
the unit. 
Above and below this central vacuum tube, the components are fixed to the body 
with copper knife-edge (CF) seals.
When the stator drives the shaft there is a reactive force of up to 50\,N.
As the stator is not fixed to the vacuum-tube this load is not 
transferred to the vacuum-tube welds. 
To prevent the coils within the stator moving, or the stator moving as a whole, 
the stator unit is clamped down inside the core body using three stacks of disc 
springs. 
These springs exert a clamping force of around 100\,N onto the stator unit when 
the upper flange of the core body is fully tightened onto the core tube. 

\subsection{Core structure}
%
The core structure of the target is provided by two robust stainless steel 
(1.4404 / 316L) components, the core tube body and the upper flange. 
These contain the stator unit and the bearings.
The other components of the target assembly, such as the electrical wiring
and the covers, are also fastened to these core components.

\begin{figure}
  \begin{center}
\includegraphics[height=8cm]{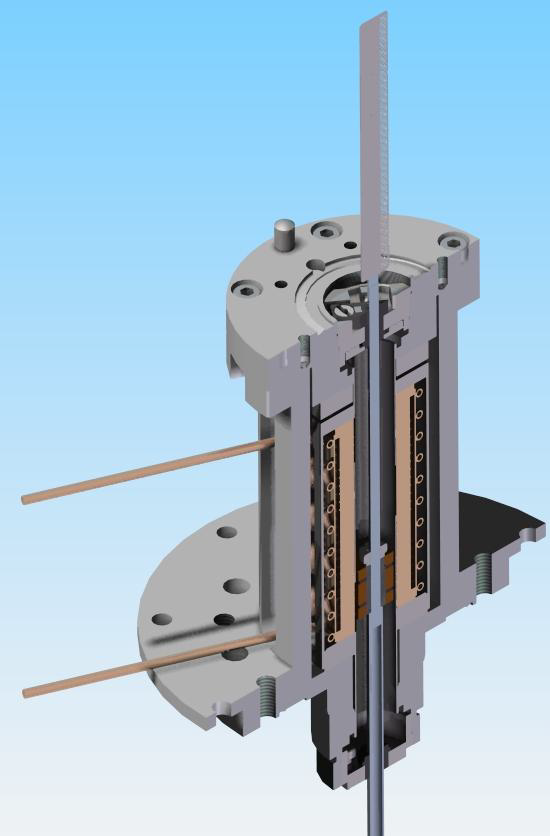}
\includegraphics[height=8cm]{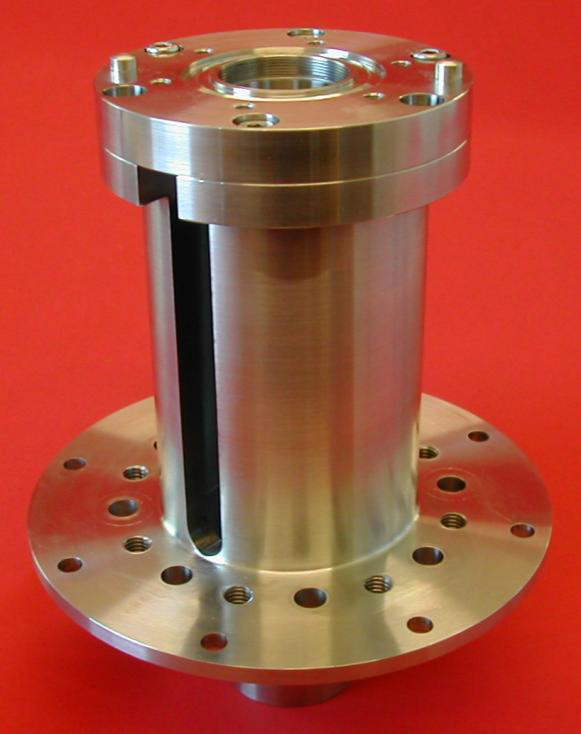}
  \end{center}
  \caption{
    (left) Core body assembly with welded vacuum tube, (right) core tube  body 
    with upper flange assembled on top with fasteners and taper dowels.
  }
  \label{Fig:Mech:Core}
\end{figure}

The core is required to align the other components accurately and stably, in 
particular to provide the accurate coaxial alignment of the permanent magnets on the
shaft and the stator coils.
Careful consideration of the interfaces between the permanent magnet and the 
stator coils was required. 
The order of the interfaces relating to this critical alignment of the 
permanent magnet and stator coils are represented by `$\rightarrow$', 
and are shown below:

Permanent magnet $\rightarrow$ Shaft $\rightarrow$ Bearings $\rightarrow$ 
Core structure $\rightarrow$ Vacuum tube $\rightarrow$ Stator coils.

This number of interfaces requires each component to be manufactured to tight 
tolerances, then accurately assembled with respect to each other. 
Coaxial alignment between the permanent magnets and the magnetic field of the 
stator reduces or eliminates off-axis torque or lateral forces on the 
magnets as they pass through the stator's changing magnetic field. 

Accurate alignment of the bearings is also very important. 
The bearing alignment, bearing-bore accuracy, straightness of the shaft and 
tolerances on the outer diameter of the shaft all build up to increase the minimum 
radial clearance tolerable between the bearings and the shaft. 
Keeping this radial clearance to a minimum ensures the shaft's travel is well 
constrained. 
If the radial clearances have no added compensation for manufacturing 
variations, the fit could be too tight which may lead to flexing of the shaft 
and possible fatigue failure or high power loading on the stator. 
To achieve accurate bearing alignment the core structure and the way it 
supports and locates the stator assembly has been simplified to the minimum 
number of components and interfaces. 
There are only two components and three interfaces between the 2 bearings:

Bearing $\rightarrow$ Core lower $\rightarrow$ Core upper $\rightarrow$ Bearing. 

Minimising the components and the interfaces is a practical solution for machining 
as there are few tight tolerances to be achieved. 
To further aid the accuracy of alignment of the bearings and reduce the 
manufacturer's liability in achieving tight tolerances, the two core parts are 
made with a slight clearance fit; they are then assembled, inspected, adjusted 
to achieve a tight coaxial alignment, then dowelled together. 
The Taylor Hobson Talyrond is used to undertake the inspection of the alignment 
of the core body components. 
The bearing seats are aligned to each other within 20\,$\mu$m 
of run-out; the repeatability of reassembly alignment with the taper dowels 
is around 10\,$\mu$m.

The bobbin that supports the stator has a central bore where the permanent 
magnet runs up and down.  
The bobbin is welded into the core assembly so that it forms a hermetic seal 
from the flange on the upper flange to the flange on the core tube body.  
These flanges incorporate CF-type seal features which use knife edges to compress 
copper seals. 
A hermetically sealed internal vacuum volume is created when the optical 
housing and bellows are fitted to the core assembly and the bellows are then 
fitted to the vessel in the ISIS ring or the test vessel. 

The core tube body has a small flange which is fitted with a larger flange 
extension.
This extension is the main reference component for positioning the 
assembly on ISIS. 
Rods are suspended from a frame, either above the beam-line or on a 
test vessel in a separate building (for offline running and trials), as described
in the next section. 
These rods fix to the larger flange extension and support the target unit for 
connection to the beam-line or vessel. 
The two-part flange is a practical solution for manufacturing as it minimises 
the material cut away from stock billets to form the components.

\begin{figure}
 \begin{center}
\includegraphics[height=8cm]{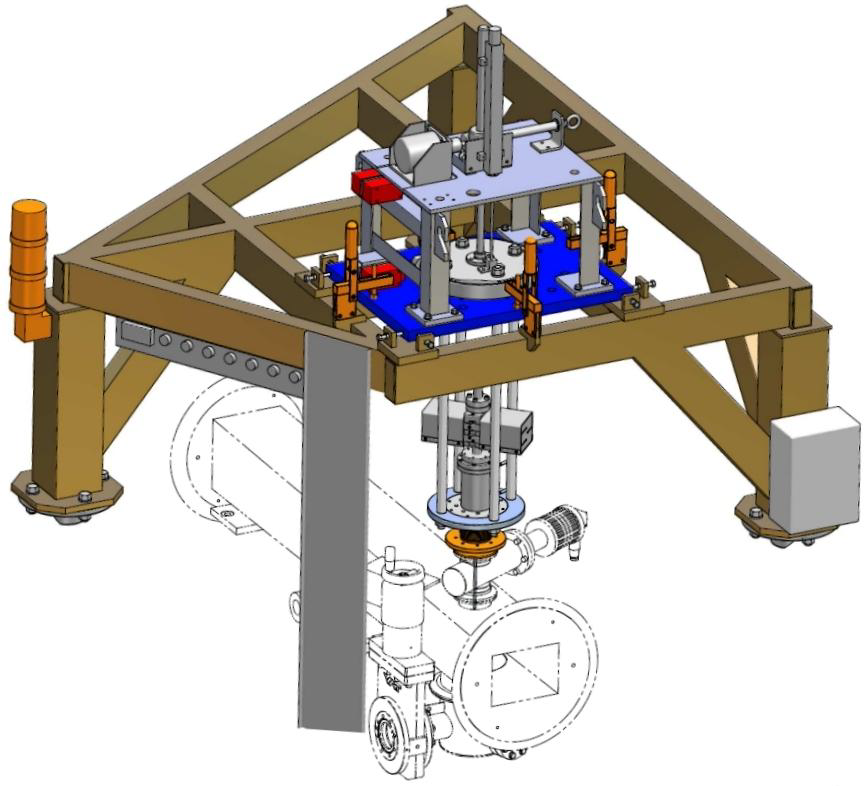}
  \end{center}
  \caption{
    Target assembly suspended by rods from motorised stage above ISIS beam
    line, connected to beam-line with bellows.
  }
  \label{Fig:Mech:Assembly}
\end{figure}

%
\subsection{Mechanical integration}
\label{Sect:Mechanical:Integration}
%
The target mechanism must be rigidly supported in a manner which
minimises the transfer of vibrations caused by the linear motor to the
synchrotron itself.
The target must also be removed from ISIS when it is not in use
and complete isolation must be possible in the case of a fault developing
in the mechanism.
These objectives are met by suspending the target from a heavy, rigid steel
frame which is itself bolted to pillars resting on the synchrotron floor, see figure 
\ref{Fig:Mech:Assembly}.
A motorised platform raises and lowers the target drive, which is connected
to the beam-pipe via a set of extensible bellows and a remotely operable
valve which can separate the vacuum systems of the synchrotron and the 
target drive.
A duplicate section of beam-pipe, complete with target support frame,
is situated in an assembly hall where all target mechanisms were commissioned.

\subsubsection{Motorised platform}
\begin{figure}
 \begin{center}
\includegraphics[height=8cm]{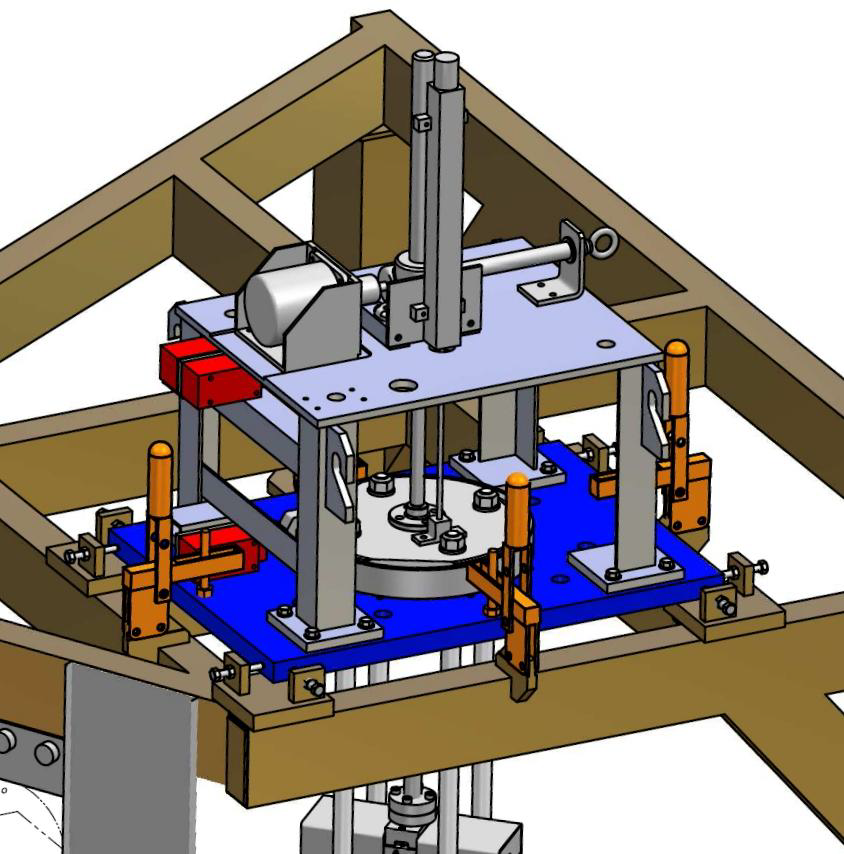}
  \end{center}
  \caption{
    Motorised platform.
  }
  \label{Fig:Mech:Platform}
\end{figure}

%
A heavy steel plate rests on the main support frame.
This carries a motorised screw-jack driven by a stepper motor (see figure
\ref{Fig:Mech:Platform}).
The jack can raise and lower a steel ring, which carries the main target
mechanism via three sturdy rods passing through holes in the plate.
The jack has a travel of 207\,mm and, when in its lowest position, the ring fits
into a locating socket.
The motor is controlled remotely, and limit-switches indicate when the 
mechanism is at its highest and lowest positions.
The switches are interlocked to the control system to prevent the equipment from being
driven outside safe limits.
An independent position switch is linked to the Personnel Protection System,
and prevents 
the target mechanism being lowered while access is allowed to the MICE Hall and
ISIS is operating.

The plate is supported on levelling screws, and its lateral position is
adjusted with locking screws to ensure the target is centred over the port
in the beam-pipe.
Once positioned, the plate is clamped in place.

During normal target operation, the support frame is moved to its lowest
position.
The tip of the shaft is then outside the beam envelope only while it is 
magnetically levitated.
When the target is not in use, the frame is raised to its highest position
and the target is well clear of the beam even if powered down.

\subsubsection{Bellows and ISIS vessel interface}
\begin{figure}
  \begin{center}
\includegraphics[height=8cm]{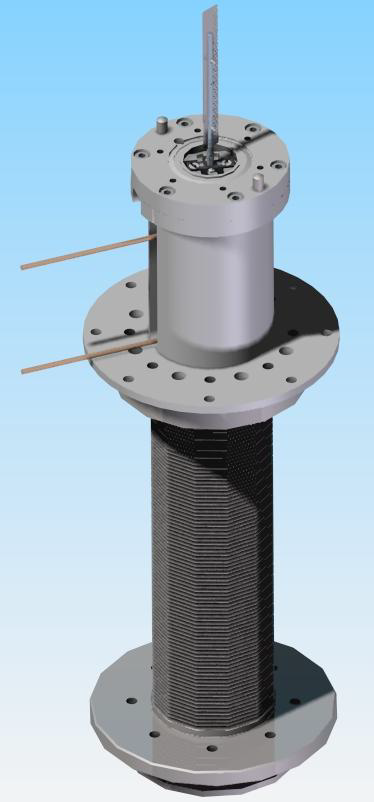}
\includegraphics[height=8cm]{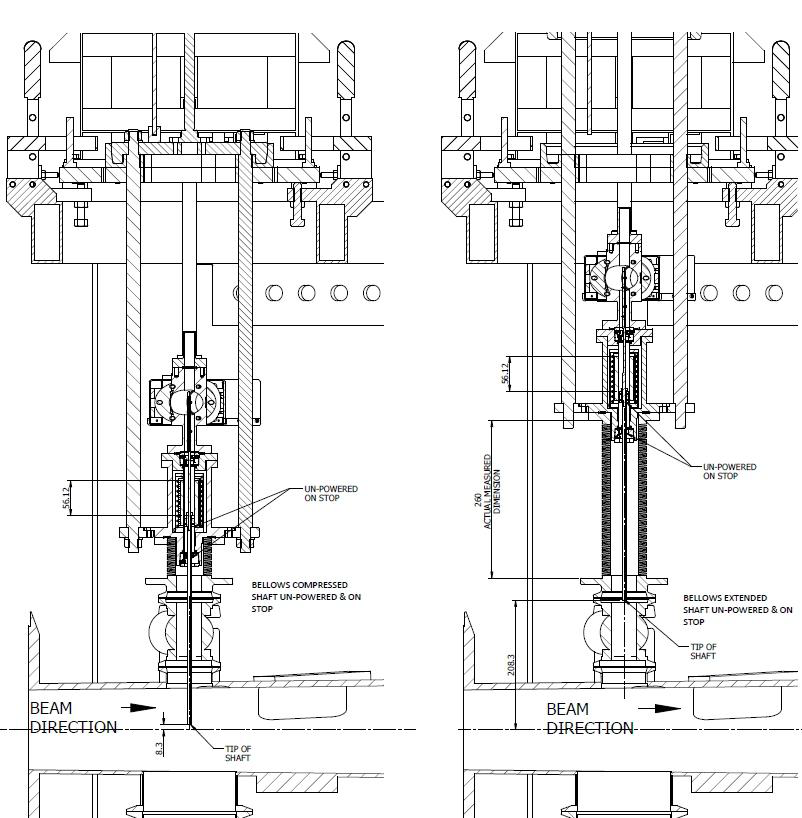}
  \end{center}
  \caption{
    (left) Bellows shown assembled under core tube body; 
    (right) Compressed and uncompressed bellows states.
  }
  \label{Fig:Mech:Bellows}
\end{figure}

%
To allow vertical movement of the target drive while maintaining a vacuum seal,
the lower flange of the drive is connected to the beam-pipe using 
edge-welded bellows (see figure \ref{Fig:Mech:Bellows}).
These have an internal diameter of 46\,mm, a compressed length of 56\,mm
between flanges and an extended length of 260\,mm.
The UHV seal between the bellows and target flange is maintained with a standard
Conflat seal and copper gaskets.

As a fault in the target mechanism could compromise the ISIS vacuum, it is
important that the target vacuum space can be isolated from the accelerator.
This is achieved with a gate valve, mounted between the flange on the upper 
face of the vacuum vessel and that at the lower end of the bellows.
Seals are formed using aluminium gaskets compressed by V-band clamps.
The valve is operated by a low-pressure nitrogen line, controlled remotely, 
and has an aperture of 40\,mm to allow the passage of the shaft.
The control system ensures that the frame cannot be lowered while
the valve is closed and that the valve cannot be closed unless the frame
is at its upper limit-switch.

The ISIS vessel is adapted for the presence of the MICE target in several
ways.
As well as the port on the upper surface to accept the target mechanism via
the gate valve and bellows, it has a thin-walled steel window to allow
the passage of pions, produced by interactions in the target, into the MICE
beam-line.
In addition, it has a glass inspection window directly below the target.
This allows visual observation of the shaft tip.

\subsection{Magnetic inspection}
%
%
To characterise the magnetic performance of the target mechanism, magnetic 
measurements are carried out on the stator and on the permanent magnet assembly. 
The equipment used is a three-axis Hall probe mounted on the end of a travelling arm.
The position of the probe can be set to a resolution of 1\,$\mu$m in three axes.

\subsubsection{Stator}

For the stator, measurements are required to define the position of the magnetic
centre of the coils relative to the mechanical centre. 
If the coils are offset, the small radial field they generate 
may have the effect of moving the shaft laterally as it passes through the stator. 
Accurate alignment of the magnetic centre is therefore important to minimise 
wear on the bearings during operation.

The coils are powered using a constant current of 7\,A; this is lower than the 
peak current when the shaft is dipped into the beam due to heating concerns -- a 
constant current of 60\,A would overheat the coils. 
The current of 7\,A heats the stator to around 50\degC when water cooling is used.

The stator is set up and aligned to the measurement bench. 
A cross-hair on the end of the Hall probe is viewed through a telescope as the 
probe is moved up and down the bench; this allows the telescope axis to be set
parallel to the bench axis. 
Cylindrical inserts with a precisely machined central hole are placed in either end 
of the stator bore in turn, to allow the stator axis to be aligned parallel to 
the bench axis. 
The estimated accuracy of this procedure is around 100\,$\mu$m at each end, which gives 
an overall alignment accuracy of 0.8\,mrad over the 180\,mm length of the stator.

\begin{figure}
  \begin{center}
\includegraphics[height=6.5cm]{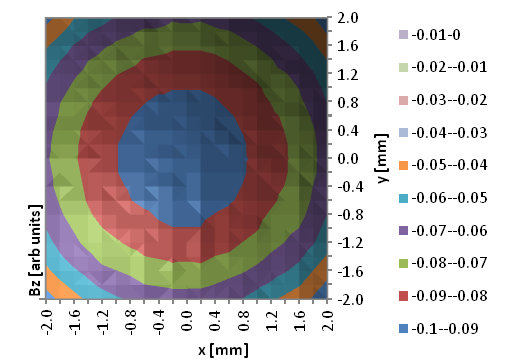}
\includegraphics[height=6.5cm]{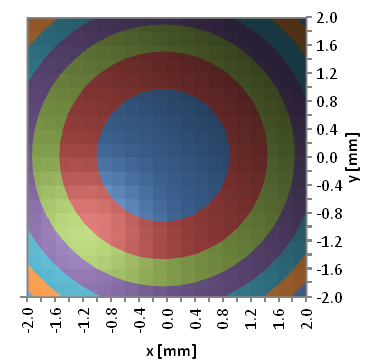}
  \end{center}
  \caption{
    Contour plot of longitudinal field ($B_z$) versus $x$ and $y$.
    The left plot shows the measured field, and the right plot
    shows a quadratic fit to the data.
  }
  \label{Fig:Mech:MagQA1}
\end{figure}

An initial scan of the stator along the longitudinal ($z$) axis shows a series of 
peaks corresponding to the individual coils. 
To find the centre of each coil, a two-dimensional scan at each peak position is 
carried out (Figure \ref{Fig:Mech:MagQA1}). 
The magnetic centre is the point at which the longitudinal field is a minimum. 
The longitudinal field is fitted to the function:
$$B_z = ax^2 + bx + cy^2 + dy + exy + f$$
The centre point can then be found from the fit to the data,
using the formulae:
$$x_0 = \frac{2bc-de}{4ac-e^2} ~;~~ y_0 = \frac{be-2ad}{e^2-4ac}$$

\subsubsection{Permanent magnets}

For the permanent magnets, measurements are carried out to determine azimuthal
variability of the radial field (i.e. $B_{r}$ vs $\theta$).
Due to variations in the PM blocks, there is a certain amount of variation in the
peak (and trough) values of the radial field.
An assembly with smaller differences between the peaks would be preferable, as it 
may reduce lateral movement and vibration of the target shaft during operation.

\begin{figure}
  \begin{center}
\includegraphics[height=5cm]{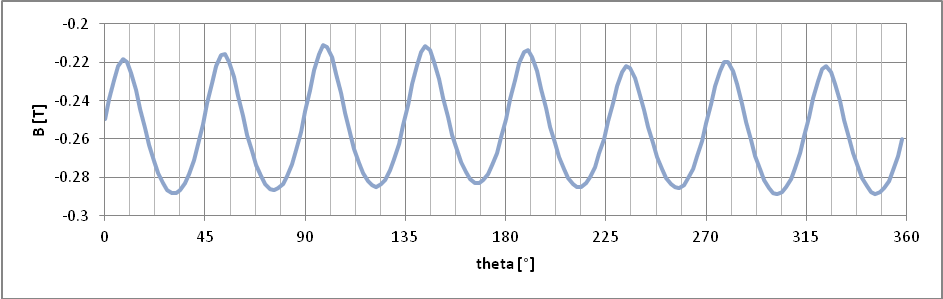}
  \end{center}
  \caption{
    Radial field versus angle for the permanent magnet assembly.
  }
  \label{Fig:Mech:MagQA2}
\end{figure}

The PM assembly is attached to a rotational stage in order to map the radial field 
as a function of angle and longitudinal position.
In this case, the Hall probe is attached to a spring-loaded head to keep it the same
distance from the PMs (about 1\,mm) as the assembly is rotated.
This was found to be necessary since the PM assembly was not parallel to the Hall
probe axis.  
The effect is small -- a few tens of microns as the PM assembly is rotated -- but 
enough to give a significant difference in the measured field, as the field drops
off rapidly with radius.
Figure \ref{Fig:Mech:MagQA2} shows a plot of the radial field versus angle.

\section{Optical position-measurement system}
\label{Sect:OpticalPosition}
%
%
In order to switch the current through the coil stack at the correct time
to drive the shaft, it is necessary to sense the position of
the shaft while it is moving.
To avoid disrupting the motion of the low-mass shaft by mechanical
contact, and to remove the necessity for electrical feed-throughs traversing
the vacuum wall, an optical method is adopted.
Furthermore, since active electrical components would not survive in the
high radiation environment near the target, optical fibres are used to convey
the signals, with all sensitive electronics being placed outside
the accelerator vault (in the experiment control room some 100~metres away).

A graticule, or vane, attached to the shaft and interrupting a light beam will
generate a series of pulses, which could be used to determine the speed but not
the direction of movement of the shaft.
If it is arranged such that two beams are interrupted by the vane, with the pulse
trains being $90^\circ$ out of phase, then both speed and direction can be 
determined.
(This is known as a quadrature system.)
With a third beam producing a pulse at only one well-defined point, a
zero of position can be defined.
In this way, by counting pulses away from the zero, an absolute position
measurement can be made.

In addition to providing feedback to the controller, position measurement
allows the trajectory of the target to be monitored and recorded by the
data acquisition system.
This enables the long-term monitoring of the stability of the mechanism 
and the diagnosis of fault conditions.

\subsection{Optical vane}
\label{Sect:OpticalPosition:OpticalVane}

\begin{figure}[!htb]
  \begin{center}
    \includegraphics[height=3.5cm]{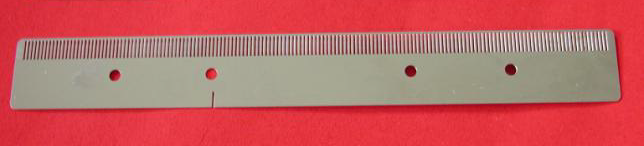}
  \end{center}
\caption{The optical vane.  
One side has 157 apertures to provide a quadrature signal.
The other has a single slot to give a fixed reference point.} 
\label{Fig:Opt:Vane}
\end{figure}

A photograph of the vane is shown in figure~\ref{Fig:Opt:Vane}.
The vane is a double-sided graticule manufactured from 0.2\,mm
thick steel, having 157 slots 0.3\,mm wide and 3\,mm long on one side of a
6\,mm wide spine. 
The whole vane is 93.9\,mm long.
The spacing between slots is also 0.3\,mm.
There is a single similar aperture two-thirds of the way down the vane on the
other side.
To protect the fine features of the vane, it has a continuous edge.
The vane was produced by photographic etching, and is attached to the
shaft by four M1.6 screws.
In order to ensure that the readout system provides reliable position
information, the manufacturing tolerances of the apertures were kept to less 
than 5\% of the 150\,$\mu$m resolution, and the tracking error along the length 
of the vane was also less than this percentage.
The vane and fixings have a mass of about 1\,g, which is balanced about the 
axis of the shaft.

\subsection{Laser source}
\label{Sect:OpticalPosition:LaserSource}

Three fibre-coupled solid-state lasers provide the light beams
which intercept the vane.
Commercial red (635\,nm) lasers, with a variable 0 to 2.5\,mW output power are used.
Visible light has advantages both for alignment and safety.
The milli-Watt power level is required due to significant losses in the optical
system owing to the number of optical interfaces.
Maximising the light on the optical sensors 
(Section \ref{Sect:OpticalPosition:OpticalSensors}) increases the signal to noise
ratio and simplifies the design of the electronic amplifiers.

In practice, the lasers are not operated at maximum power, but at about 1\,mW.
This leaves sufficient overhead (both in light source and amplification gain)
to adjust for any degradation in the optical fibres, e.g.\ increased
attenuation as a result of radiation damage.

\subsection{Optical fibres}
\label{Sect:OpticalPosition:OpticalFibres}

Two types of optical fibre are used.
On the transmitting side, single mode fibres are required to achieve
the necessary small spot size at the focal point and hence at the 
plane of the vane.
On the receiving side, 200\,$\mu$m multimode fibres are used.

The single mode fibres used are SM600 with FC-to-FC connectors~\cite{Thorlabs1}.
These fibres have a core of pure silica, which is a radiation-hard material and so 
ideal for the operating environment of the target.
If single-mode fibres were used on the receiving side, the collimators would 
have to be aligned to an extremely high precision and it would be hard to
achieve an adequate light transmission.
Multimode fibres have a higher acceptance, so make the alignment less critical.
The fibres used are BFH37/200 purchased from Thorlabs, 
with SMA to SMA connectors~\cite{Thorlabs2}.

\subsection{Collimators, lenses and mechanical mount}
\label{Sect:OpticalPosition:Collimators}

\begin{figure}[!htb]
  \begin{center}
    \includegraphics[height=8cm]{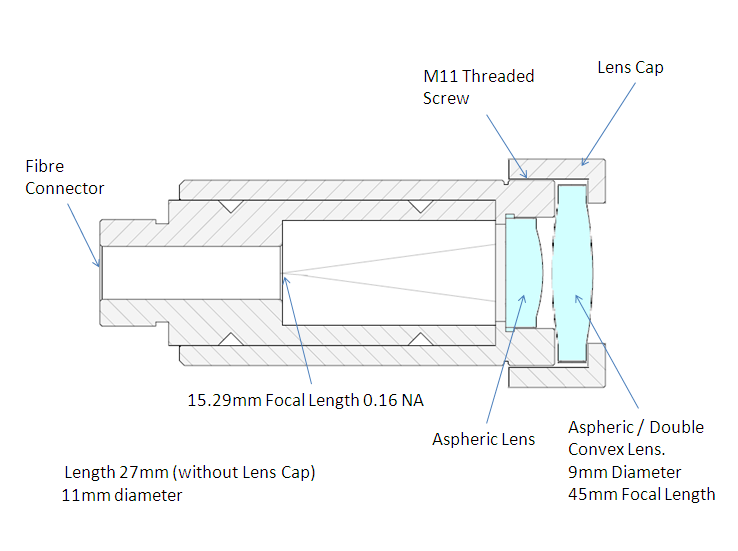}
  \end{center}
\caption{Cross section through a collimator and lens cap.} 
\label{Fig:Opt:Collimator}
\end{figure}

Collimators and lenses are used to focus the light from the fibres to a
spot in the plane of the optical vane (where it needs to be significantly 
smaller than the pitch of the vane) and to receive the transmitted light
coupling it into the return fibre.
The collimator is a commercial unit which produces parallel light from
the diverging beam emitted by the fibre.
Collimators with a focal length of 15.3\,mm are used, as the longer focal
length minimises the final spot size.
The focusing lens, with focal length 45\,mm,
is attached to the front of the collimators by inserting
it into a holder which is screwed onto the front of the collimator
as shown in figure~\ref{Fig:Opt:Collimator}.
On the transmitting side, aspheric double achromatic lenses are used to obtain 
the minimum spot size.
On the collecting side, where focusing is less critical, double convex lenses
are used to re-collimate the beam.
The lenses are MgF$_2$ coated to minimise reflections and maximise the light 
transmitted into the fibres.

\begin{figure}[!htb]
  \begin{center}
    \includegraphics[height=8cm]{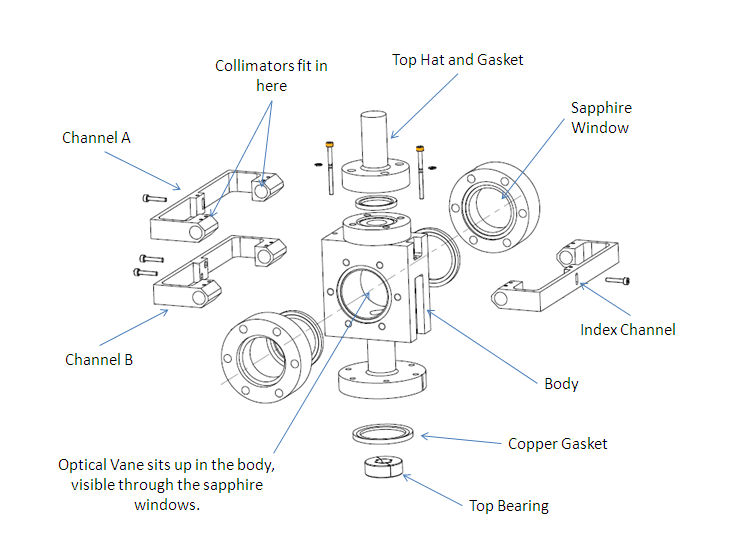}
  \end{center}
\caption{An exploded view of the optical mount.
The optical vane sits inside the main body, where the collimators and lenses 
are able to focus the light.}
\label{Fig:Opt:OpticalMount}
\end{figure}

The optical vane moves inside the accelerator vacuum system, while all 
optical elements are kept outside the vacuum chamber to allow alignment
adjustments.
A mechanical mount was produced which has two circular sapphire windows,
to allow entry and exit of the light and to provide a rigid assembly
to support the optical collimator and lens units.
This is illustrated in figure~\ref{Fig:Opt:OpticalMount}.
The flat windows are bonded into metal flanges and are rated for use in
UHV environments.
The pair of collimators for each optical channel is held in an arm that
wraps around the mount.
The exact alignment of each collimator is performed by adjusting pairs of 
opposed grub screws, with 4 pairs per collimator: horizontal and vertical at
front and back of each collimator.
The ``Channel A'' and ``Index'' arms are also adjustable in the vertical 
direction to allow the correct quadrature phase to be obtained and an
appropriate index position to be set.
The mount has proved to be easy to set up and extremely robust.
Once aligned on the bench, no further optical adjustments have been 
necessary during periods of operation lasting several years.

\subsection{Optical sensors}
\label{Sect:OpticalPosition:OpticalSensors}

The multimode fibres are returned to an SMA photodiode (H3R880IR)\cite{Farnell}.
This is a broad spectrum photodiode (400 to 1100\,nm).
As the maximum velocity of the target is less than 10\,m\,s$^{-1}$, the maximum 
data rate per channel is only 33\,kHz, well within the response capability of 
these devices.
The outputs from the photodiodes are amplified and conditioned before being
converted into digital signals.

\section{Stator operation and power electronics}
\label{Sect:PowerElectronics}
%
%
\subsection{Introduction}

The target drive is a three-phase, brush-less, permanent-magnet DC linear 
motor.
Before considering how the force on the magnetic assembly changes
with respect to its position within a powered coil stack, it is necessary
to understand how the coils in the stator are wired and how they are
switched. 

\begin{figure}[!htb]
\begin{center}
    \includegraphics[height=8cm]{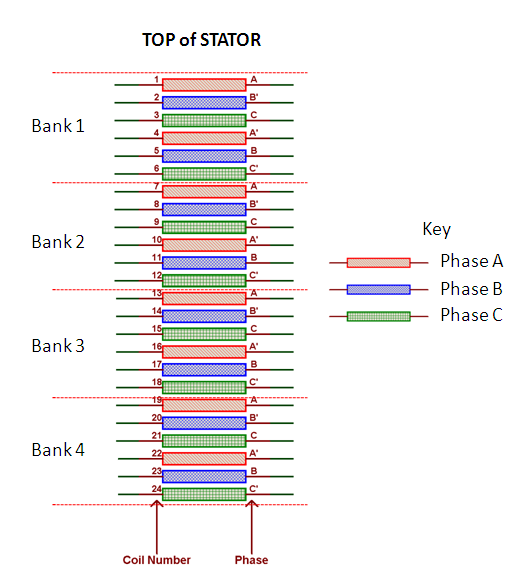}
\caption{The relationship of phase of the coils to their physical
position within the stator. Coils marked with a prime are wired so
that current flows through them in the opposite direction to
those that are unprimed.} \label{Fig:Power:COrder1}
\end{center}
\end{figure}

\begin{figure}[!htb]
\begin{center}
    \includegraphics[height=8cm]{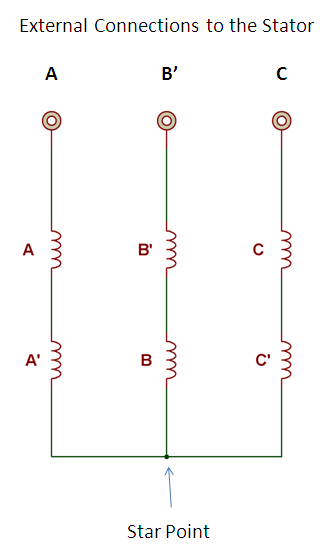}
\caption{The wiring of the coils in the stator,
illustrating the star point and the three external
connections from the stator. Each coil in the diagram represents 4
coils i.e. the $A$ coil represents the $A$ coil from banks 1 to 4.}
\label{Fig:Power:SPoint1}
\end{center}
\end{figure}

The 24 coils in the stator are split into three
phases $A$, $B$ and $C$, with eight coils in each phase.
Starting from the top of the coil stack (see figure \ref{Fig:Power:COrder1}),
the coils are lettered in a cyclic
sequence that follows the pattern, $A$ $B'$ $C$ $A'$ $B$ $C'$. 
A block of six coils labelled in this way is called a ``bank''. This
sequence is repeated four times so there are four banks of
coils in the target stator, as shown in figure \ref{Fig:Power:COrder1}.
All the $A$ and $A'$ coils are wired together in series, as are
the $B$ and $B'$ coils and the $C$ and $C'$ coils. 
The unprimed and primed coils are wired such
that when a current
passes through either an $A$, $B$ or $C$ coil in a clockwise direction, the same
current passes through an $A'$, $B'$ or $C'$ coil in an anticlockwise direction.
The induced
magnetic field direction for a primed coil is therefore opposite to that
of an unprimed coil.
There are two connections for each phase, one at the top of the stack
and one at the bottom. 
The three separate connections at the bottom of the
stack are wired together to form the ``star-point'', while
the three at the top are connected to the stator power supply, as
shown if figure \ref{Fig:Power:SPoint1}.  
If current is fed into one of the phases then it must
return through one (or both) of the other phases. 
Therefore
when current flows, at a minimum two phases must have current passing through them.

\subsection{Coil current switching sequence}

\begin{figure}[!htb]
\begin{center}
    \includegraphics[height=8cm]{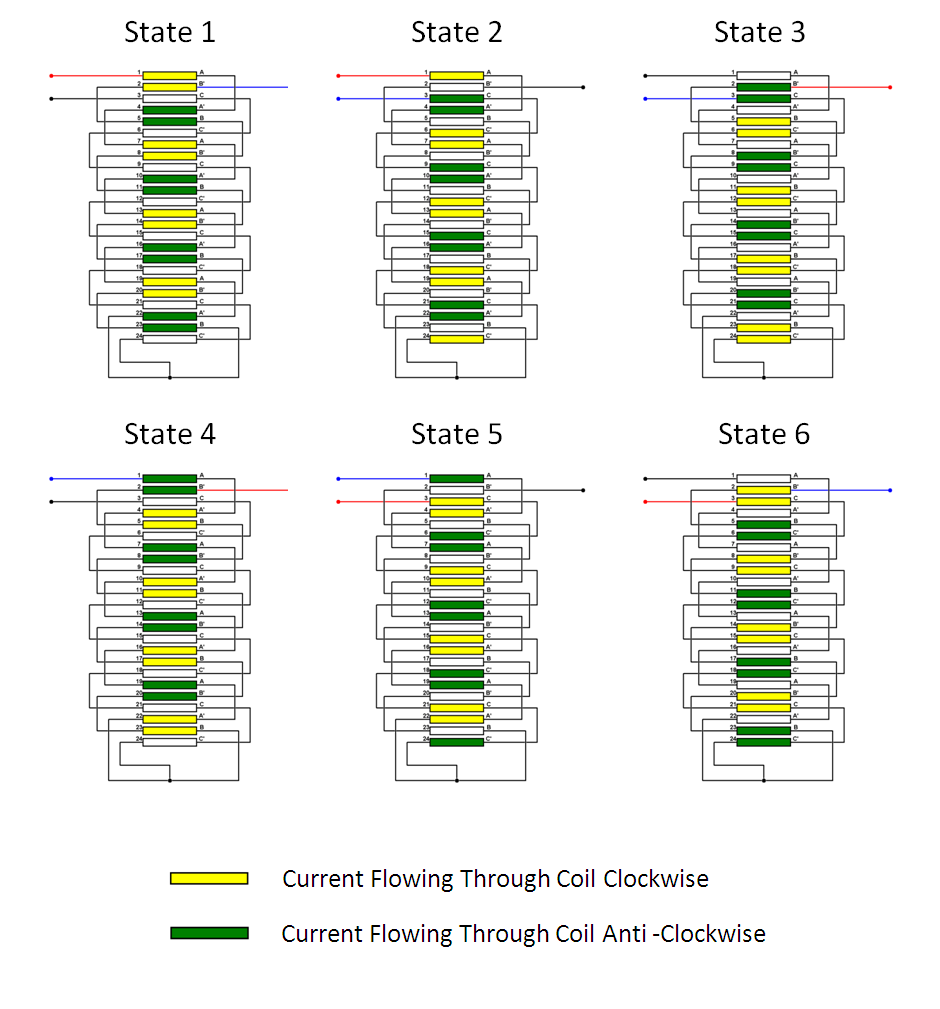}
\caption{Switching the coils through the circular sequence given
by States 1 to 6 gives an apparent motion in the coil switching
upwards through the stator as illustrated. Reversing the
sequence gives an apparent downwards motion. The pattern repeats
itself every six coils. The permanent magnets on the shaft
can be made to interact with the fields produced by passing
sequentially through these states to produce motion of the
shaft.} \label{Fig:Power:CoilSeq1}
\end{center}
\end{figure}

For the target mechanism, only two of three phases are ever powered
at the same time\footnote{We have recently changed to powering three phases simultaneously.}, 
i.e.\ one phase is switched to provide the current source and another
phase provides the return path. 
This operation is slightly different from the usual method of driving each phase
with a phasor, but reduces complexity at the cost of only a small reduction in
efficiency.
With three phases, current can therefore
be switched through the stator in six different ways.
Using the external connection labels $A$, $B'$ and $C$ as shown in 
figure \ref{Fig:Power:SPoint1},
these six states are:
$A{\rightarrow} B'$, $A{\rightarrow} C$, $B'{\rightarrow}A$, $B'{\rightarrow} C$, $C{\rightarrow} A$ and $C{\rightarrow} B'$.
By stepping through these six states in a predetermined order,
the coils can be switched to create a ``ripple''
motion in the magnetic fields generated within the stator,
as illustrated in figure \ref{Fig:Power:CoilSeq1}.
The direction of motion is determined by the direction in which
these states are stepped through; 
when reversing this order the ripple motion
moves in the opposite direction.
By tracking the position of the magnets and correlating this position
to a given state, target motion can be achieved. The required order of the states
to observe this apparent motion upwards through the stator is shown in
table \ref{Tab:SixStates1}.

\begin{table}[hbtp]
\begin{center}
\caption{The six states that create an apparent upward motion in the
fields generated within the stator.  The states are circular. Moving
through the states in the opposite direction reverses the apparent
motion.
}
\label{Tab:SixStates1}
\begin{tabular}{| c | c |}
\hline State & Current Flow\\
\hline \hline 1 & $A{\rightarrow} B'$ \\
\hline 2 & $A{\rightarrow} C$\\
\hline 3 & $B'{\rightarrow} C$\\
\hline 4 & $B'{\rightarrow} A$\\
\hline 5 & $C{\rightarrow} A$\\
\hline 6 & $C{\rightarrow} B'$\\
\hline
\end{tabular}
\end{center}
\end{table}

\subsection{Magnetic assembly and modelling}

As described in section \ref{Sect:LinearMotor:PermMag}, the magnet
assembly is composed of 3 radially magnetised rings, with a total
length of 18\,mm, matching the depth of 6 coils.
This distance corresponds to the symmetry of the axial field, and alignment 
of the magnets with
these fields gives the ability, with suitable feedback, to either
hold the magnetic assembly in a predetermined position or to maximise
the accelerating force on the permanent magnets.
 
\begin{figure}[!htb]
\begin{center}
    \includegraphics[height=8cm]{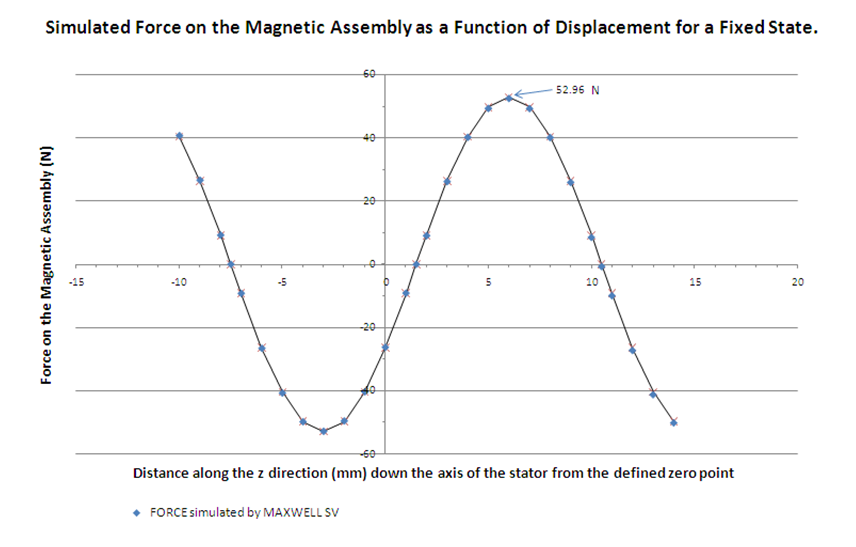}
\caption{Simulation of the force on the
magnetic assembly as a function of its position within the
stator when the stator is in one of its six states with a coil
current of 58\,A. A good sinusoidal fit can be made to these points
as illustrated.} \label{Fig:Power:ForceCurve1}
\end{center}
\end{figure}

A simulation of
the interaction between the magnetic assembly and the coils was produced using 
a magnetostatic model in Opera 3D{\cite{MAXWELL:SV}}.
This simulates the stator in ``State 4''. 
(The state is arbitrary due to the symmetry of the device.) 
The model is axisymmetric in \emph{R, z} and the origin is placed at the 
centre of the stator, corresponding to a plane that sits on top of the 13$^{th}$ coil. 
For this model a current of 58\,A was assumed to flow through the coils, approximately
equal to the peak current during operation.
The simulation ran over several iterations and, with each iteration,
the magnetic assembly was moved in either 0.5\,mm or 1\,mm
increments axially along the centre line of the stator
over a total distance of 20\,mm. 
The maximum force on the
magnetic assembly was calculated for each position.
The results are shown in figure \ref{Fig:Power:ForceCurve1},
which shows a clear
sinusoidal pattern for the resultant force on the magnetic assembly
and a sine wave can be fitted cleanly to the simulation results.
Although not illustrated here, the sinusoidal pattern of the
force curve repeats every 18\,mm, the length of one
bank of coils.
Further simulations confirmed that as the 
magnets move towards the end of the stator bore there is no
deviation from this sinusoidal pattern, as the magnet
assembly is always completely contained within the coil stack so 
end effects are minimal.

\subsection{Zero-force points}
\label{Sect:PowerElectronics:ZeroForce}

From figure \ref{Fig:Power:ForceCurve1} it can be seen that there are two points 
for each repetition of the
sine wave where the force on the magnetic assembly is zero.
One of these zero-force points is unstable; movement of the
magnetic assembly away from this point results in a force that
pushes the magnetic assembly further away. 
However the other zero force point is stable, and movement of the magnets 
away from this
zero-force point results in a restoring force; effectively, this point
is the centre of a magnetic well.
The stable zero-force points provide a mechanism by which the shaft can be
levitated passively.  
The shaft will sit at an equilibrium position where the force of gravity
is counteracted by the restoring force exerted by the magnetic
force. 
Any small deviation from its equilibrium position would
result in the shaft undergoing damped harmonic motion (where the damping
would primarily come from the frictional forces between the shaft and
the bearings).
The drive is designed to be a high-acceleration device,
accelerating the shaft in excess of 80\,$g$,
so providing a force greater than $80 m g$, where $m$ is the mass of the 
shaft complete with magnets etc. 
Stable levitation occurs when the electromagnetic force on the 
shaft balances $m g$. 
To a good approximation the force scales linearly with the coil
current so stable levitation of the shaft in close proximity
to these zero-force points can be achieved with a current that is much smaller
that that required to achieve the high accelerations needed to insert the
shaft tip into the ISIS beam.

\begin{figure}[!htb]
\begin{center}
    \includegraphics[height=8cm]{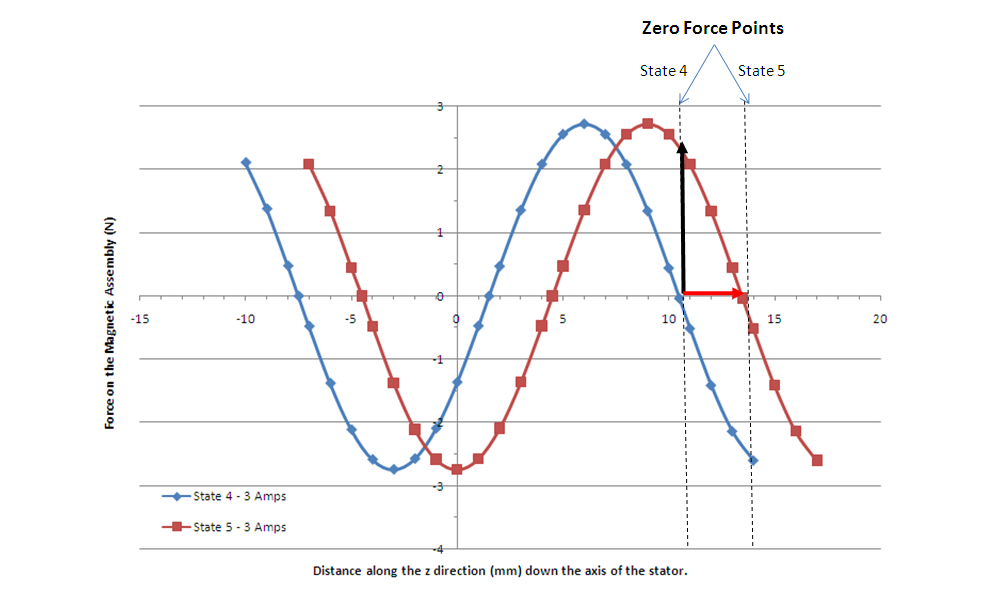}
\caption{If
the shaft was sitting at the zero-force point that corresponds to
``State 4'' and the stator was switched to ``State 5'' then the
magnets would see a restoring force (black arrow) that would move
the shaft to the new zero force position (red arrow) 3\,mm further up
the stator. By repeatedly incrementing or decrementing the states
the shaft can be moved up and down the stator body. This simulation shows
the force for a 3\,A coil current typical of that used to levitate
the shaft.} \label{Fig:Power:ForceCurve2}
\end{center}
\end{figure}

The zero-force points in figure \ref{Fig:Power:ForceCurve1} are shown for
the stator in ``State 4''. If the state sequence is progressed
then the positions of these zero-force points move in step with the
state sequence. Moving forward through the sequence moves the zero
force points up through the stator in 3\,mm increments, likewise moving
backwards through the sequence moves the zero force point down the stator
in 3\,mm decrements. If the zero-force point is used to levitate
the shaft then the target will track the movement of
these zero-force points as the state sequence is progressed. When
the state changes the shaft will move to the new point,
as a restoring force will push the shaft to the new equilibrium
position. This is illustrated in figure \ref{Fig:Power:ForceCurve2}. The shaft
will undergo damped harmonic motion as it settles at this new point.

The ability to move and hold the shaft in this way is
utilised to control the target shaft position when the target
is not actuating.  For example by switching the coils to the appropriate state when the
target system is powered up, the shaft is picked up from its
resting position, also known as its ``parked'' position, and
then moved to its holding position by
progressing cyclically through these states. This final state, the
``hold mode'', then holds the shaft indefinitely until actuation is required. 
Reversing the process allows the
shaft to be lowered back down to its ``park-mode''.


Using the stator in this way to control the shaft limits
the ``hold position'' of the shaft to one of a set of predefined points
that are 3\,mm apart. This means that the hold position is not
entirely arbitrary, but the 3\,mm step size gives enough freedom
to ensure that the shaft is held out of the ISIS beam.
This system of moving the target shaft by allowing the magnets to
track the position of the zero-force points is entirely passive and
does not require any positional feedback to operate. 

\subsection{Actuation}

\begin{figure}[!htb]
\begin{center}
    \includegraphics[height=16cm]{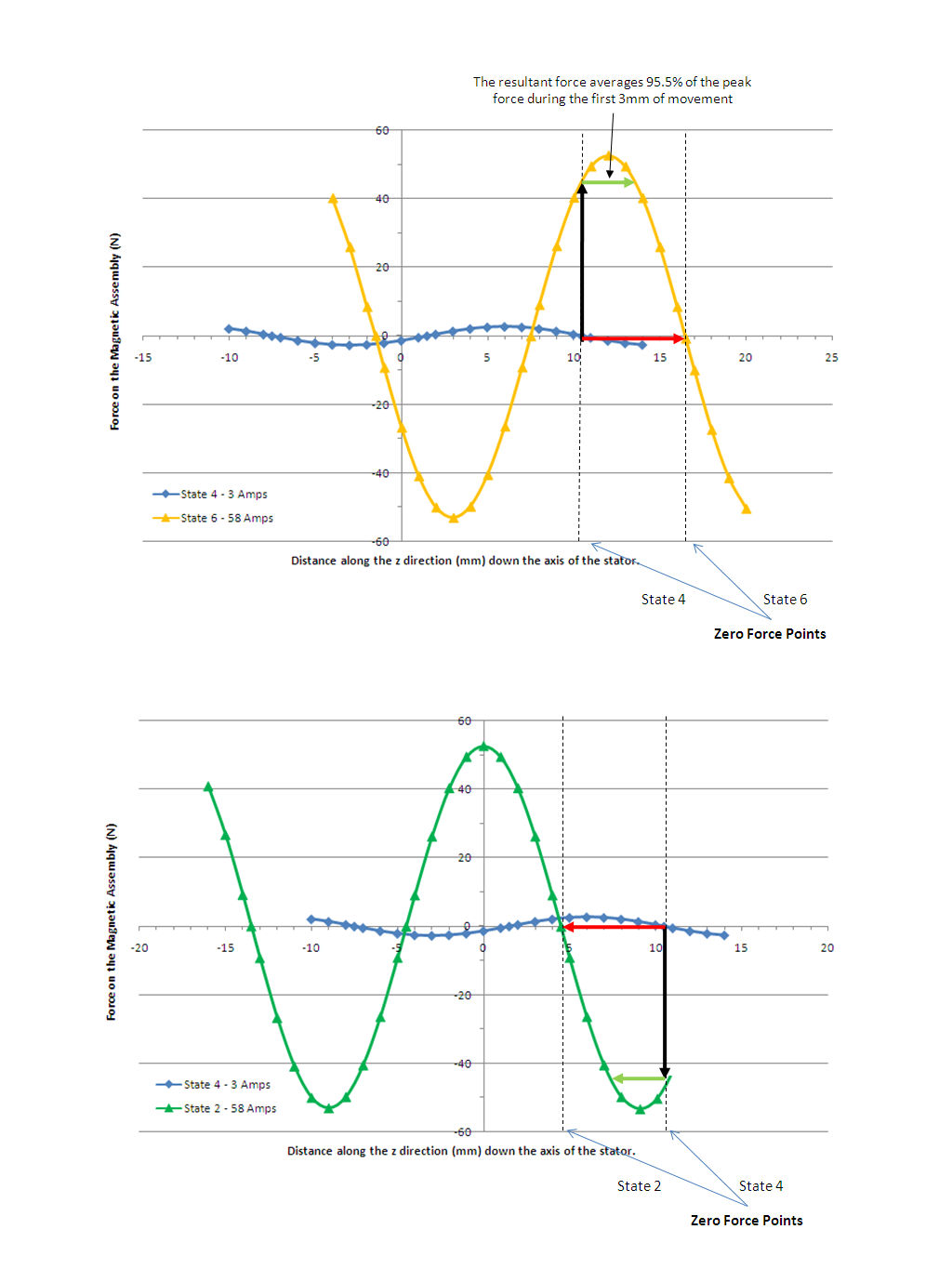}
\caption{If
the shaft was sitting at the zero-force point that corresponds to
``State 4'' and the stator was switched to ``State 6'' (top) or
``State 2'' (bottom) then the magnets would see a restoring force
(black arrow) that would  move the shaft towards the location of
the shifted zero-force position. During the first 3\,mm of movement the shaft would
see an average force that is $\approx$96\% of the peak force.}
\label{Fig:Power:ForceCurve3}
\end{center}
\end{figure}

From figure \ref{Fig:Power:ForceCurve1} it can be seen that the
zero-force point is half way between two maximum-force points;
we define a maximum-force point as a position within the coil stack where the magnets on the
shaft would experience a maximal repulsive force. 
For each set of two maximum-force points, one of them will push the magnets in 
one direction whilst
the other will push the magnets in the opposite
direction.  Figure \ref{Fig:Power:ForceCurve1} shows that
these maximum-force points are positioned $\pm$4.5\,mm away
from the zero-force points; this is of course true for any zero-force point in 
any one of the six possible states.
It can be seen that the forces change
very little $\pm$1.5\,mm either side of the peak.
Integration of the fitted sine wave $\pm$1.5\,mm either side of the peak shows
that the average force is 95.5\% of the peak force. If the shaft
was levitated at a zero-force point and the coil state was
either to increment or to decrement by two states this would put the
magnets 1.5\,mm on the far side of one of the maximum-force points as the zero-force
point would have been moved by 6\,mm. The shaft would then
accelerate back towards the shifted zero-force point.
In either case the resultant force would cause the shaft to pass through the 
location of the maximum-force point during the first 3\,mm of acceleration,
as illustrated in figure \ref{Fig:Power:ForceCurve3}.

\begin{figure}[!htb]
\begin{center}
    \includegraphics[height=9cm]{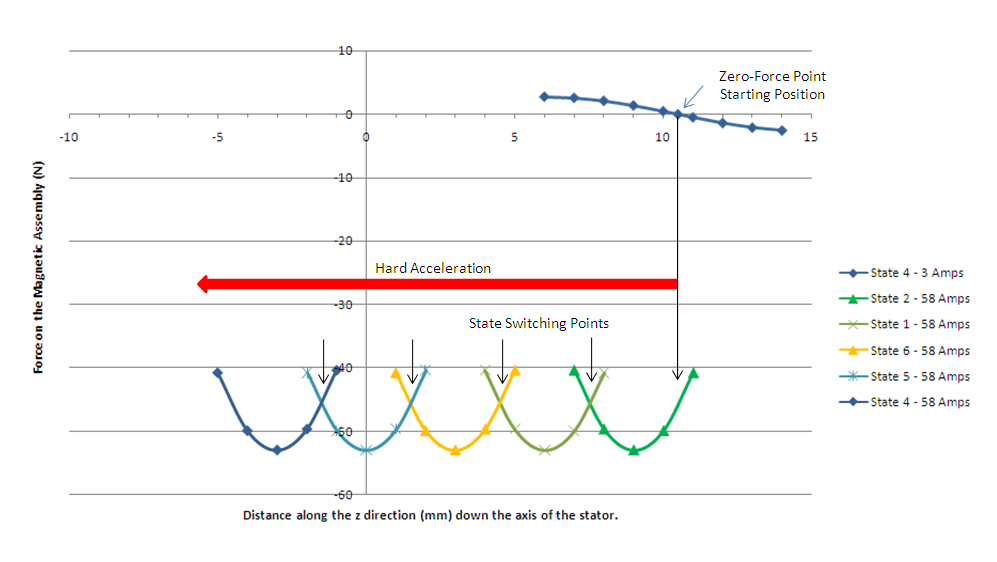}
\caption{The
shaft moves from levitation (3\,A coil current) to acceleration (58\,A
coil current). By tracking the position of the magnets the states
can be switched to ensure the shaft continually accelerates using
the maximum available force. This results in a large accelerating force on
the shaft.} \label{Fig:Power:ForceCurve4}
\end{center}
\end{figure}

If, after switching the coil two states, no further switching was done the shaft would
continue to accelerate towards the new zero force point 
6\,mm further up or down the coil stack and would execute
damped harmonic motion about this point until it came to rest.
However because the position of the
shaft can be tracked using the optical quadrature system, another
state change can be made when the target has travelled 3\,mm, placing
the magnets 1.5\,mm on the far side of the next maximum-force point.
This has the effect of maintaining a maximal accelerating force on the shaft.
This process can be continued down the entire length of the coil stack as
illustrated in figure \ref{Fig:Power:ForceCurve4}.
If the shaft is accelerated via this mechanism, deceleration can be achieved by
switching the coil state by three positions. This has the
effect of placing a force of equal magnitude but opposite direction on the permanent
magnets. Once again, by referring to figure \ref{Fig:Power:ForceCurve1} and
comparing this to figure \ref{Fig:Power:CoilSeq1}, it can be seen that a
switch of three states changes the direction of the accelerating force
because it simply reverses the current flow through the coils.
If a previously accelerated shaft is decelerated by this process
then there will be a point where the shaft will change its
direction of motion. 

By monitoring the sense in which the position reading
is changing, the quadrature system is able to determine the
direction of motion of the shaft.
A reversal of direction can be used as a trigger to
capture the shaft at the nearest zero-force point at the end of
an actuation. This is done by
switching to a state that places a zero-force point
close to the magnet's current position.
The coil pitch of 3\,mm dictates that the shaft will never be
more than 1.5\,mm away from a possible zero-force point and so,
providing the
shaft does not have a high velocity, capture of the shaft at the
zero-force point is inevitable.

\begin{figure}[!htb]
\begin{center}
    \includegraphics[height=8cm]{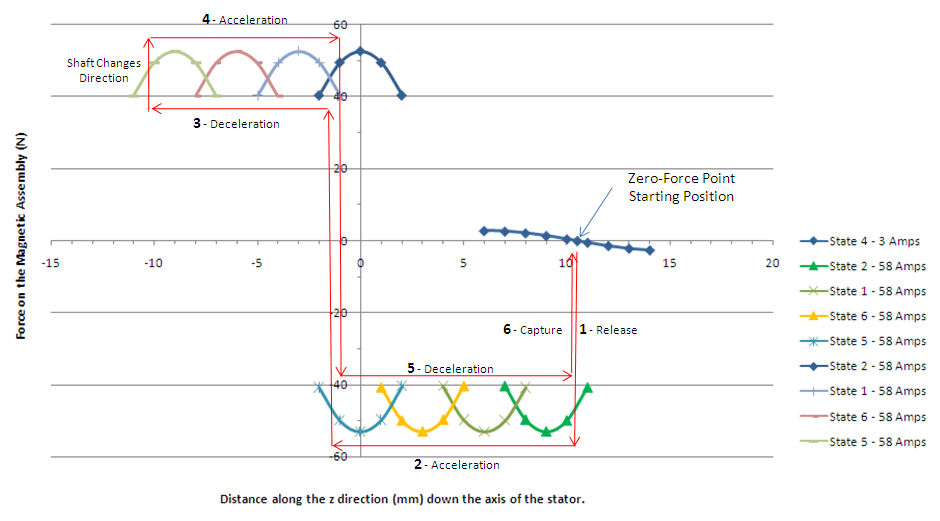}
\caption{An
actuation. The shaft is accelerated downwards, decelerated,
accelerated upwards, decelerated and then captured. The position of
the shaft must be tracked accurately so the controller knows into
which state to put the coils to maximise the
accelerating/decelerating forces.} \label{Fig:Power:ForceCurve5}
\end{center}
\end{figure}

The processes just described provide the necessary mechanism 
to accelerate the target into and
out of the ISIS beam whilst enabling its capture again at the end of
the cycle.
Controlling the motion in
such a manner is called ``actuating''.
The minimum positional resolution required to control the
shaft in the way described is 1.5\,mm whereas the optical system
provides the position to within 150\,\microns\!\!. As will be
described in the next section, the high resolution of the optical system
allows better control of the stroke of the
shaft.
The use by the control electronics of the mechanisms described above 
is given in section \ref{Sect:Controls},
while figure \ref{Fig:Power:ForceCurve5} shows the principle employed.

\subsection{Coil switching and current control}

It has been shown that by wiring
the stator in three phases and applying the appropriate currents
to those phases it is possible to control the movement of the shaft. 
To do this effectively there are two principal design requirements.
The first of these is the ability to switch current bi-directionally through
any two of the three phases, and the second is the
ability to control the amount of current that passes through the coils.

\begin{figure}[!htb]
\begin{center}
    \includegraphics[height=8cm]{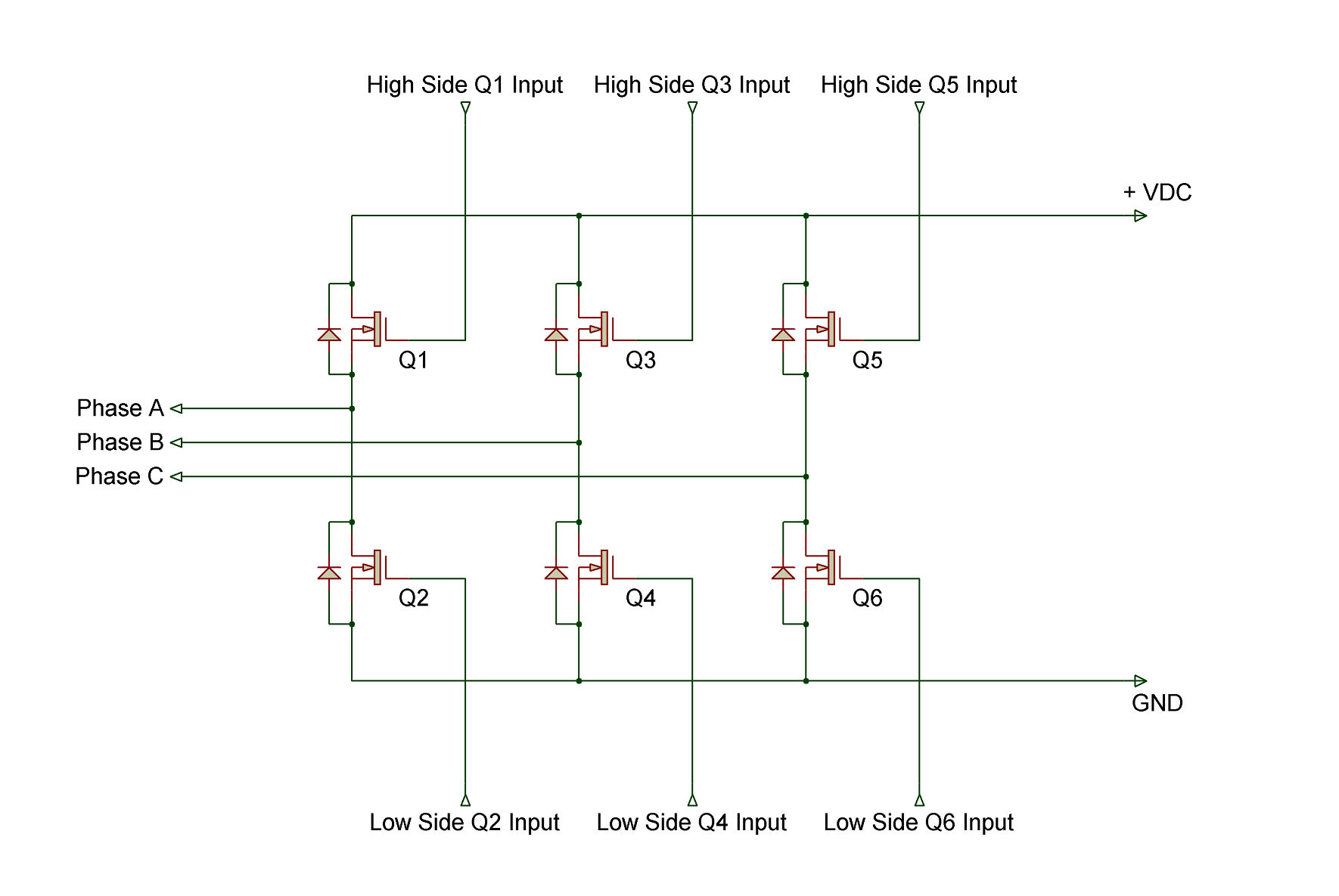}
\caption{The Hex
Bridge.  This circuit allows bidirectional current control through
any two of the three phases of the coil stack by application of
appropriate control signals to the six inputs that switch the
transistors on.} \label{Fig:Power:HexBridge1}
\end{center}
\end{figure}

Bi-directional switching of the three phases of the
coil stack can be achieved using six transistors arranged in three pairs, 
where each pair of transistors is
connected together in series between the power rails of the power
supply. The mid-point of each pair of transistors
then connects to one of the three phase wires of the coil stack. Figure
\ref{Fig:Power:HexBridge1} illustrates how these transistors are connected.
This type of circuit is known as a ``Hex Bridge'' or ``Three
Phase Inverter''.
The three transistors across the top of the circuit are called the
``high-side transistors'' as they are connected to the positive power
rail, while the other three are the ``low-side transistors''.
It is possible to see how this circuit can be used to switch the
current bidirectionally through the phases by comparing table
\ref{Tab:SixStates2} with figure \ref{Fig:Power:HexBridge1}. 
The inputs are labelled Q1--Q6 in the figure.

\begin{table}[hbtp]
\begin{center}
\caption{How the six stator states can be switched
using the Hex Bridge.}
\label{Tab:SixStates2}
\begin{tabular}{| c | c | c |}
\hline State & Current Flow & Hex Bridge Inputs\\
\hline \hline 1 & $A{\rightarrow} B'$  & Q1 and Q4\\
\hline 2 & $A{\rightarrow} C$ & Q1 and Q6\\
\hline 3 & $B'{\rightarrow} C$ & Q3 and Q6\\
\hline 4 & $B'{\rightarrow} A$& Q3 and Q2\\
\hline 5 & $C{\rightarrow} A$ & Q5 and Q2\\
\hline 6 & $C{\rightarrow} B'$ & Q5 and Q4\\
\hline
\end{tabular}
\end{center}
\end{table}

One risk that needs to be managed with this circuit is that if both transistors 
connected to a given phase are turned on
simultaneously, e.g.\ if inputs Q1 and Q2 are activated simultaneously,
then a short circuit is created between the power supply rails.
This will permanently damage the transistors due to
excessive current flow. This type of short
circuit due to transistor switching errors is known as a
``shoot-through''.
Shoot-through is also possible if account is not taken of
the time that it takes to switch transistors on and off.
It is therefore necessary to ensure that the second transistor is only
turned on after an appropriate delay.
Shoot-through is not normally a
problem when stepping through the state sequence in a circular
fashion as there is always at least one state on a given phase between a
low-side transistor turning off and the complementary high-side transistor
being turned on (or vice versa). 
The minimum delay requirement is satisfied since the frequency of state
changes is no more than a few kHz.
However this is no longer true when the force
direction is suddenly reversed.
In this case both pairs of transistors experience a change in state
and so a suitable delay must be imposed
by the controller.


When the shaft is being levitated in its holding position the
gravitational force will pull the shaft slightly away from the
zero-force point such that there will be a small restoring force
acting upon the shaft which exactly counters gravity.
The amount of counter-force that the stator needs to supply to the
shaft is small, reflecting the low mass of the shaft.
Ideally this equilibrium position needs to be as close to the zero-force 
point as possible although a small offset can be accounted for.
It can be seen from figures \ref{Fig:Power:ForceCurve2} and 
\ref{Fig:Power:ForceCurve2} that
$\frac{dF}{dz}$ is maximal at the zero-force point so the amount of
displacement required along the \emph{z} axis to produce the
restoring force is minimal and approximately proportional to the applied
current over short displacements.

The amount of current passing through the coils needs to be large enough that
the levitation is stable but small enough that power dissipation is minimised.
A current of around 3\,A has been found to
satisfy these criteria. This can be seen from figure
\ref{Fig:Power:ForceCurve3}; the shaft, magnets and vane have a mass of $\sim$57\,g
so the gravitational force is about 0.56\,N. This force is counteracted by
the electromagnetic force at a displacement of $\sim$500\,{\microns}
from the true zero-force point.

\subsection{The target power supply}

During actuation the shaft will be accelerated at $\sim$80\,$g$.
It has
been found that a coil current of $\sim$60\,A is needed to obtain
the required acceleration. This current, if sustained, would quickly
overheat and damage the coils and so care must be taken to ensure
that the actuation current is only supplied to the coil stack for the
required amount of time. 
By pulse-width modulating the current through the coil stack,
the average current that flows for
a given duty cycle is, to a first approximation, linearly
proportional to the maximum current that would flow through the
device if the duty cycle were set to 100\%. 
To ensure that enough overhead was built in to permit future 
upgrades, the power supply was specified to operate at up to
300\,V and to provide a current of up to 100\,A.
The coil stack only requires the
high currents for relatively short periods of time and the
average current drawn is significantly lower than the peak current.
It is therefore more economical and efficient to power
the driving circuit from a capacitor bank which can
provide the short, high-current pulses on demand. The
capacitor bank is charged by an external power supply,
designed to provide both the necessary charging current to
top the capacitor bank up and the holding current.
This capacitor charging unit or CCU
effectively provides the average current that the stator uses, whilst
the capacitor bank is there to provide the peak currents when
required. The size of the capacitor bank attached to the hex-bridge
is 70\,mF and is rated to 400\,VDC.

The use of a capacitor bank is inherently safer than a linear
power supply because, should a fault occur that leaves the system in
a state that demands high current, once the capacitor bank has
discharged the current is limited to that provided by the CCU.  
The CCU used only provides a current of 4\,A, and into a 3.6\,$\Omega$ load 
(the DC resistance of two phases of the stator at 20\degC\!\!) this
gives a power dissipation of only $\sim$60\,W. 
This is significantly lower than the power
dissipation during normal operation when actuating at
1\,Hz and so limits the energy deposition into the coil stack. 
At 70\,mF the capacitor bank stores a significant amount of energy. 
This energy could be deposited into either the coil stack or the
bridge circuit under a fault condition. 
In order to reduce the voltage between any part of the coil stack and 
ground, a split supply providing $-115$\,V and +115\,V is used (rather than 
0\,V and +230\,V).
This necessitates the use of two separate capacitor banks.

To switch the high currents necessary to drive the target
drive, Integrated Gate Bipolar Transistors (IGBTs) are used.
These transistors
switch current quickly and saturate very close to the
power-supply
rail voltage due to a small internal resistance
($R_{\rm{ON}}$). Both of these characteristics are good for
minimising power dissipation within the transistors
making them very efficient at transferring power to the
load. By using these transistors a circuit that can switch a
significant amount of power can be built with minimum footprint.
The IGBTs that were chosen for the target power supply are power
devices and can
switch up to 250\,A.
All six sections of the gate drivers (one for each IGBT) are taken
to a floating DC power supply as this gives better
protection to the individual circuits.
The logic circuitry that drives the gates on the IGBTs is
supplied with power from a separate supply that is integral to
the unit. 
Each of the two capacitor charging units uses a Xantrex 300-4, which
can supply up to 4\,A at up to 300\,V.

Snubber circuitry has been added to the power supply
to damp the high frequency transients created by the
pulsed switching of the transistors. This also helps to stabilise
the current flow through the motor and reduces electromagnetic
noise emitted by the motor, its power supply cable and the driving
circuitry.

\subsection{System placement in ISIS}\label{CH:POWER:PLACEMENT}

The power
supply cannot be close to the target mechanism as the radiation
produced by ISIS would be likely to cause premature failure of the electronics.
It is also advantageous to have the electronics accessible so that
power supply maintenance can be performed
without having to wait for the synchrotron to be opened. The high
currents required by the target mechanism and its low impedance 
mean that the distance to the
power supply should be minimised to reduce ohmic losses in the power
cable connecting the two together. 
For this reason the power supply has been installed on
the outside wall of the synchrotron at a location as close to the
target area as possible, requiring a
cable length of 25\,m.
The control electronics for the target system are
situated in the MICE Local Control Room, a
significant distance from both the target and the power electronics.
The distance between the control and power electronics is 70\,m and
that between the control electronics and the target drive is 100\,m.

\subsection{Fibre-optic links}

The fibre-optic cables that run between the control electronics
and the drive's optical block have been discussed in
section \ref{Sect:OpticalPosition}. 
Optical fibres are also used
to transmit the six signals between the controller and the power supply. 
The use of
optical fibres guarantees signal integrity, completely isolates the
power supply from the controller and eliminates the risk of noise on
the signal lines which could cause additional problems with
either the controller or the transistor drivers in the power supply.

The fibre-optic link between the control and power electronics uses 
a commercially available optical
transmitter/receiver pair. The transmitter is a high-power
infra-red LED with a bandwidth up to 5\,MHz. This is
sufficient as a 1\% duty cycle resolution on a 20\,kHz PWM
signal requires a bandwidth of $\sim$2\,MHz. 
The receiver is a stand-alone unit that gives a TTL-compatible output for 
use in the hex-bridge driving circuitry.

The optical fibres used are the same multimode fibres used to
return the laser light in the quadrature optical counter (BFH37-200). 
These fibres are well matched optically to the
transmitters, and experiments in the lab showed lower losses using
these cables than using the standard polymer cables that came
with the transmitter/receiver pairs. 

\section{Target control}
\label{Sect:Controls}
\begin{figure}[!htb]
\begin{center}
  \includegraphics[height=12cm]{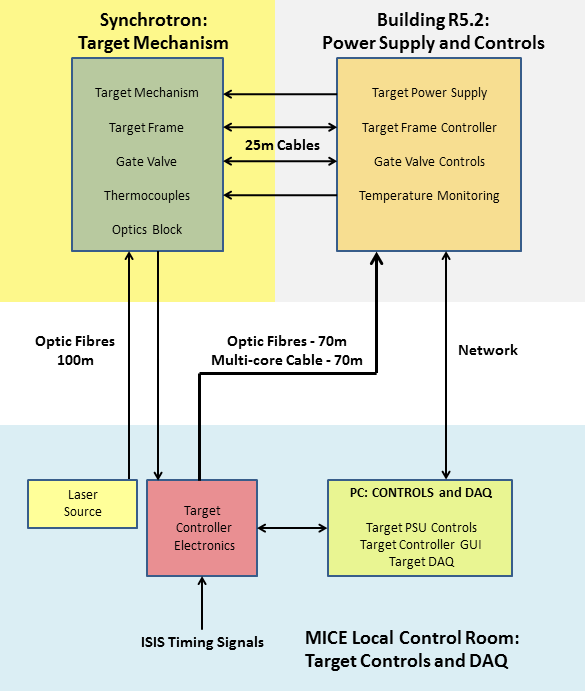}
\caption{The major components of the target
mechanism installation at ISIS, showing the relationship between the components.} 
\label{Fig:Controls:Install1}
\end{center}
\end{figure}

%
%
This section gives a description of how the target is controlled by the
system electronics.
The interface between the electronics and the computer that records
the performance of the target, the Data Acquisition (DAQ) system,
is also discussed. Figure \ref{Fig:Controls:Install1} gives an overview of the major
components of the target system and how they relate to each other as installed
at the Rutherford Appleton Laboratory.

\subsection{Target controller overview}

The target controller is implemented on a Field Programmable Gate Array (FPGA).
For the controller the
logical units of the FPGA are configured to create a finite state machine (FSM) that both
interfaces the controller to the outside world and controls the operation of
the target mechanism.
A Xilinx XC3S1000 \mbox{Spartan-3} FPGA is used~\cite{UG331}. 
The FPGA has been bonded to a custom PCB that provides access to most of the
FPGA's IO pins and has a USB interface. The USB interface allows the FPGA to
communicate with a PC,
permitting soft control of the underlying FSM based upon its current state
through a GUI interface.
This FPGA/PCB combination has been used in other experiments and has proven to
be reliable~\cite{FPGA_PCB}.

\begin{figure}[!htb]
\begin{center}
  \includegraphics[height=8cm]{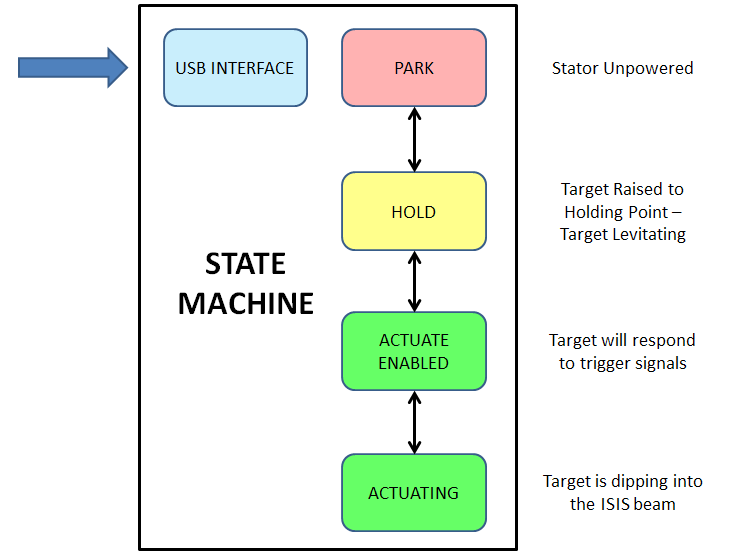}
\caption{The four fundamental states of the target system.}
\label{Fig:Controls:FourStates1}
\end{center}
\end{figure}

The four states defined for the target
mechanism are
Parked, Hold Mode, Actuate Enable and Actuating. 
These are illustrated in figure \ref{Fig:Controls:FourStates1}, with the
paths which connect them.
The job of the controller
is to manage the movement of the shaft, ensuring that it is in the
correct place for its given state, the correct state being dependent
upon input from both external signals and the user. 
The controller's FSM has many sub-states to ensure that the target operates
correctly in a safe manner; this level of detail is not
described here. 

\subsection{Control: park, hold and actuate enable modes}

%
Before the target can be actuated it must be moved into its
``hold'' mode. 
%
When the shaft is lowered, its tip will intersect with the ISIS beam
if the jacking frame is in its ``in beam'' position.
Therefore, the controller is provided with interlocks to ensure that the
shaft is in the hold position before the frame can be lowered and to cause 
the frame to be raised should the shaft move to the parked position when
the jacking frame is lowered.

An enable key-switch on the front of the controller unlocks the system and permits it
to receive control signals over the USB connection.
Providing that all the interlocks are good, 
the controller can be commanded from the GUI to raise the shaft to its hold
position.
The latter is determined by a register value that is loaded over
the USB interface and so is configurable in software. The default value is set
to raise the target by 51\,mm, enough to hold the tip
out of the ISIS beam when the frame has been lowered.
The holding position is achieved by advancing the zero-force point up the
coil stack as described in an earlier section.
Once the shaft has been raised to its hold position a
check is performed to ensure that a correct count has been
obtained  by the quadrature counter and that an index signal
has been received. An incorrect count or missing index signal could indicate a
problem with the quadrature system and actuation would not then be possible.

The change of state from hold mode to actuate-enable mode reflects a change
of the internal state of the controller which is initiated by the user from the
control PC.
However the system will not enter actuate-enable mode until 
the position counter reads an appropriate value, there are no other
internal errors and all external interlocks are satisfied.

\subsection{Control: actuation and capture}

``Actuation'' refers to the state when the stator is actively
accelerating the shaft.
Actuation is completed
when the shaft has been electromagnetically recaptured at a
zero-force point. 
Reliable capture of the target is essential to ensure it does not
fall into the beam. 

\begin{figure}[!htb]
\begin{center}
  \includegraphics[height=8cm]{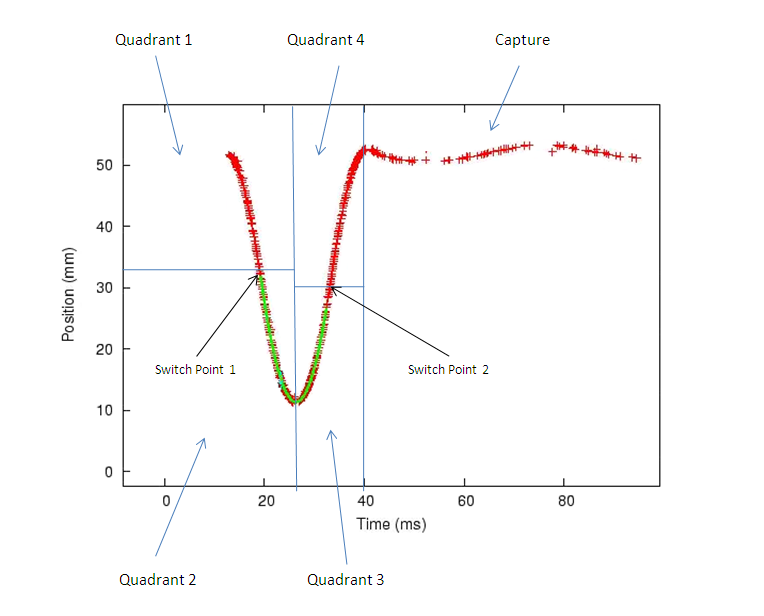}
\caption{The
actuation trajectory is split into four quadrants from the
controller's point of view.  Switch Point 1 reverses the coil
currents and so determines the actuation depth. Switch Point 2
determines the capture point. These two points are not at the same
position due to the decay on the capacitor bank.} 
\label{Fig:Controls:4Quad}
\end{center}
\end{figure}

Actuation is performed actively; the coils are switched to provide the
maximum force on the permanent magnets at all
times during the actuation process. For this force to be
maintained the coil switching has to track the
position of the shaft and an accurate measure of the its position
is essential.
The actuation process goes through four
distinct stages known as ``quadrants''.
The capture of the target is considered a separate process.
The four quadrant states are shown in figure \ref{Fig:Controls:4Quad}.

Upon receipt of a trigger signal, the control system enters
``quadrant~1'' of the actuate sequence. 
Here the controller causes the current through the coils to be switched
such that the shaft accelerates
downwards. The position of the shaft is tracked and the coils are
switched to maintain the maximal force until the shaft
reaches position ``switch-point 1''. This is the
position at which the coil currents are reversed and therefore determines the
actuation depth. 
The position of switch-point~1 is calculated by the control software
given the desired strike entered via the GUI.
The strike may be controlled to a precision of $\sim$300\,\microns since the 
resolution of the quadrature system is 150\,\microns.

When the currents are reversed
the shaft begins decelerating and the controller enters
``quadrant 2''. 
The shaft is decelerated
until it reaches the position where it changes direction as
indicated by the quadrature counter. At this point there is no physical
change to the system and the coil currents are kept the same;
however this change in the shaft direction defines the point at
which the controller state moves into ``quadrant 3''.
The shaft now accelerates upwards until it reaches the position
defined by ``switch-point 2''. 
Here the coil currents are reversed
again and the shaft begins decelerating; this is accompanied by the
controller state moving into ``quadrant 4''. The shaft continues to
decelerate and at the point where its motion
changes direction again the controller enters the capture state.
Because the shaft will have some small downward velocity by the 
time the change of direction is
detected, it is necessary to reverse the current flow again for a short period
of time, typically a few hundreds microseconds, to arrest the residual velocity. 
This ``kickback'' proved to be necessary for reliable target capture.

When the controller is in the capture state it calculates the location of 
the nearest zero-force point.
It then switches the coils to place this zero-force point as close as
possible to the shaft's current location; the shaft is then left to be
captured passively into this zero-force point. Ideally if ``switch-point 2''
is set to the correct value then the capture position should be the same as the
hold position. If the zero-force point does not correspond to the
hold position then after capture the shaft will be moved to the hold
position ready for the next actuation.

Switch-point 2 is offset with respect to switch-point 1 because the decay
on the capacitor bank means that the rate of acceleration in
quadrants one and two is greater than that in
quadrants three and four. Changing the strike changes both the
switch points by an equal amount. If the value of switch point 2 is too high or
too low, the shaft will be captured at either a higher or
lower zero-force point.
If under/over capture occurs persistently then the controller will
automatically make a correction to the switch-point 2 offset until the shaft is
being captured at the hold point. This system of quadrants to
define the actuation cycle allows the shaft to be tracked accurately by the
controller as it passes through the trajectory, ensuring reliable actuation.

\subsection{The actuate trigger signal}

The design of the actuate-trigger system is based on the following assumptions:
\begin{itemize}
\item Actuation must be synchronised with the ISIS cycle and occur only when
  the MICE apparatus is ready to take data; and
\item The time at which the target intercepts the beam must be adjustable.
\end{itemize}
For a strike of 44\,mm, the duration of the actuation cycle is approximately
30\,ms.
The shaft reaches the apex of its trajectory after 14--15\,ms, the precise timing
depending on the strike.
The combined effects of gravity and the capacitor discharge cause the shaft to 
accelerate more slowly out of the beam than it accelerates into the beam.

The tip of the shaft is required to intersect the beam for the last
$\sim$2\,ms of the acceleration cycle.
Since beam is extracted from the synchrotron 10\,ms after it is injected,
the actuation cycle must be initiated significantly before the start of
this cycle.
ISIS is able to provide a ``machine start'' (MS) signal that arrives up to 
5\,ms before the start of a cycle;
however this provides insufficient time for the target to intercept the beam at 
the end of the subsequent cycle.
A programmable delay is therefore used to trigger actuation in time to
intercept a {\it later} cycle, the number of cycles' delay being agreed in advance
between ISIS and MICE.
The ISIS operator monitors the beam loss produced by the target to ensure
that the loss occurs only on the specific spills that the target is set to intersect.
The number of cycles' delay is chosen to be $2^n$, where $n$ is typically 8.
The delay is therefore set by the controller to be 
$2^n\times 20\;\rm{ms}-15$\,ms.

To optimise particle production for MICE while keeping proton-beam
losses to acceptable levels, a fine programmable delay is used to tune the 
time within a spill at which the tip of the shaft intercepts the beam.
Fine control of the time at which the shaft reaches the apex of its trajectory
is required to compensate for the change in the trajectory of the shaft as 
a function of the strike.
The deeper the target has to dip, the earlier it must be triggered.

\section{Performance}
\label{Sect:Performance}
%
%
\subsection{Particle Production and Beam Loss}
Several signals are provided by ISIS to allow MICE to verify that beam losses
induced by operation of the target occur at the correct time and are within 
acceptable limits.  These include the sum
of the signals from the beam-loss monitors in sectors 7 (where the target is situated)
and 8 (immediately downstream), the sum of 
the signals from all the beam-loss monitors and the outputs of the
vertical and horizontal beam-position monitors closest to the position of
the target\cite{ISISBLM}.
These voltage signals are fed into a National Instruments (NI)
 6254 PCI card~\cite{NI6254} which samples
 the signals at 100\,KS/s for 50\,ms around the target actuation. The shaft
 position is also communicated to the card using a 10-bit parallel connection,
 and is recorded simultaneously with the voltage signals. The position
 is read at 200\,KS/s to allow ``deglitching'' of the asynchronously-sampled
 parallel bits. The combined signal and position data are then both
 displayed online for real-time feedback and written to disk for later
 analysis. The recording of particle data is handled by the main MICE
 DAQ, enabling later comparison to the target data by matching
 appropriate timestamps.

For optimum operation the target mechanism must maximise particle
 production for MICE whilst simultaneously minimising losses in ISIS.
 The target achieves this by chasing the shrinking proton beam and only
 intercepting the beam during the final $\sim$2\,ms of the acceleration
 cycle during which the pion-production cross section is the highest.
 Then, to prevent losses at the next ISIS injection 10\,ms later, the
 target tip must be completely outside the beam envelope by this time. 

\begin{figure}[!htb]
  \begin{center}
  \includegraphics[height=8cm]{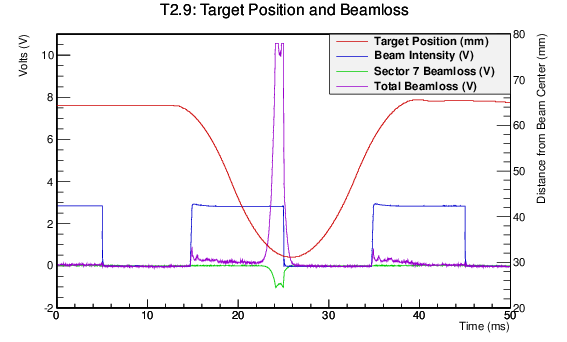}
    \caption{ Signals from ISIS showing the beam intensity, total instantaneous 
    losses and the summed instantaneous losses from sector 7, with the target 
    trajectory overlaid. Two spills are shown by the blue beam-intensity line. 
    The total instantaneous losses are a positive-going signal, while the 
    individual sectors are negative-going.  }
    \label{Fig:Performance:Actuation}
  \end{center}
\end{figure}

Figure~\ref{Fig:Performance:Actuation} shows the trajectory of the shaft
together with the proton intensity.
Two ISIS cycles are shown, the first spill being the one into which the 
tip of the shaft is inserted.
Taking the arrival of the actuate-trigger as $t = 0$, the shaft begins to
scrape the shrinking proton beam halo at $\sim$20\,ms when
small beam losses can be observed.
The tip then intercepts the
 beam at around 22\,ms and induces larger beam loss during the
 final 3\,ms. The shaft finally moves out of the beam and
 reaches a safe distance of 58\,mm before the next injection cycle at 35\,ms.
 The position of the shaft tip at maximum insertion is normally measured 
 relative to the nominal beam 
 centre and this coordinate is referred to as Beam Centre Distance (BCD); 
 as the target moves further from the beam this value increases.

\begin{figure}[!htb]
  \begin{center}
  \includegraphics[height=8cm]{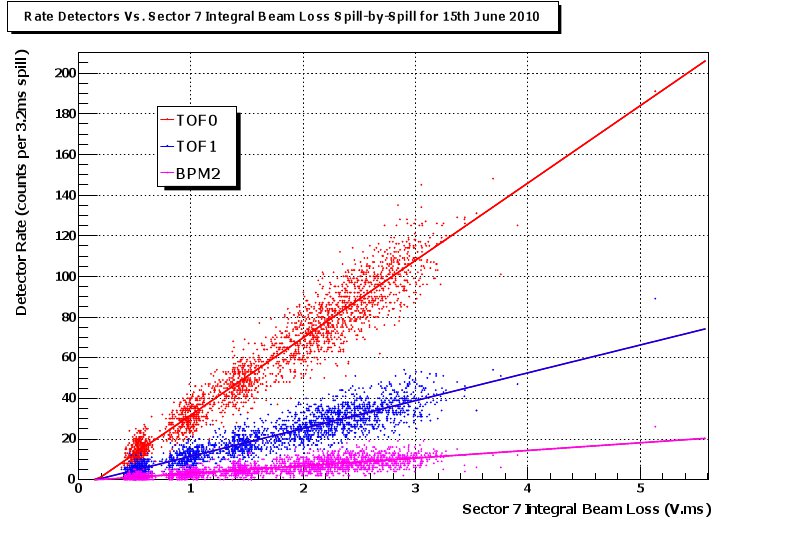}
    \caption{Particle rates in various detectors as a function of sector 7 beam loss. As the beam
    loss increases the particle rate for a given beam setting increases.
    From \cite{DOBBSTHESIS}.}
    \label{Fig:Performance:Rate}
  \end{center}
\end{figure}

The particle rate in MICE plotted as a function of the beam-loss in sector 7
integrated over the acceleration cycle is shown in figure~\ref{Fig:Performance:Rate}.
The data were collected with the MICE Muon Beam set to deliver negative muons 
with a central momentum of 238\,MeV/$c$\cite{MICE_Beam}. 
The figure indicates that the muon rate delivered to MICE depends linearly
on the beam loss generated in the synchrotron.
For a more detailed description of the dependence of particle rate on beam loss
the reader is referred to \cite{MICE_Beam} and \cite{DOBBSTHESIS}.

\begin{figure}[!htb]
  \begin{center}
  \includegraphics[height=8cm]{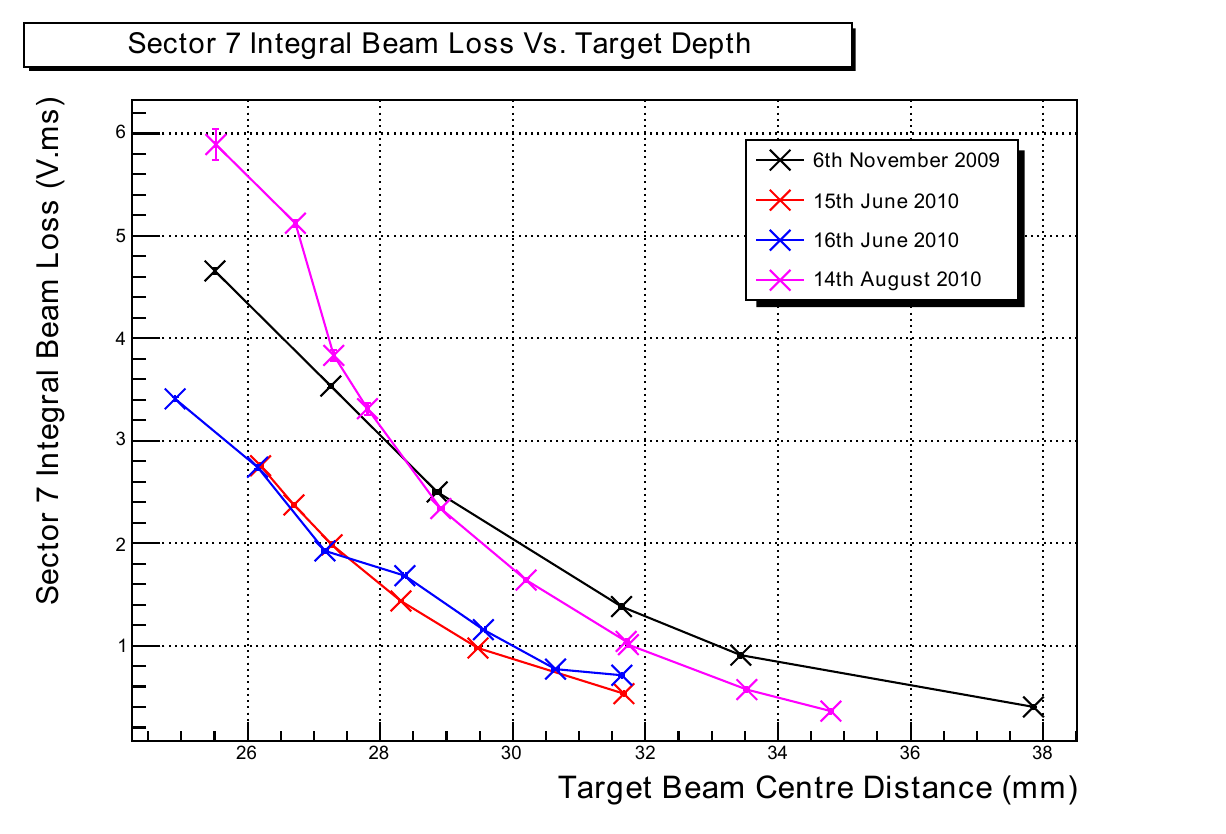}
    \caption{Beam loss as a function of target depth. While the target 
    is capable of generating in excess of 6\,V.ms of beam loss, the limits set 
    by ISIS for regular running are normally 2\,V.ms. From \cite{DOBBSTHESIS}. }
    \label{Fig:Performance:Beamloss}
  \end{center}
\end{figure}

Figure~\ref{Fig:Performance:Beamloss} shows the integrated beam-loss in 
sector 7 plotted as a function of BCD.
The minimum BCD which still yields acceptable beam-loss is $\sim$19\,mm.
At a BCD of 25\,mm the
target was capable of generating between 3 and 6\,V.ms of beam loss depending
 on ISIS conditions. 
During routine operation in 2010 and 2011, a limit of 2\,V.ms 
integrated beam-loss was imposed\cite{BLOSSLIMIT}. 
At fixed BCD, the beam-loss varies from run to run due to changes in
ISIS beam conditions.
The BCD is adjusted regularly to accommodate these fluctuations.

\subsection{Target Lifetime}
For operation on ISIS, the target must be capable of being operated
reliably for periods significantly in excess of a typical User Run, 
typically of six-weeks duration.
The system is monitored to identify the early signs of wear as a damaged or
malfunctioning target would disrupt the smooth running of ISIS and 
prevent MICE from taking data. 
Wear on the Vespel bearings limits the lifetime of the target mechanism
to millions of actuations. The bearings are buried
deep within the stator body preventing direct visual inspection and monitoring.
Therefore, indirect techniques have been developed to monitor the performance 
of the mechanism and the wear on the bearings.
 
\subsubsection{Monitoring}
Monitoring the mechanical performance of the target mechanism is performed using a
 list of key values returned from the controller after each actuation. This
 list includes data from the actuation, such as the time to reach switch-point 1,
 the minimum position reached and any errors which occurred. This information
 is ideal for long-term monitoring because it contains only key values, reducing
 the volume of data and the processing time. In addition, since the data is
 calculated by the control algorithm during actuation it is much more precise
 and flexible than that collected by the NI card. The data is analysed
 offline using a simple ROOT\cite{ROOT} script to study the variation in performance 
as a function of time.
 
\subsubsection{Bearing Performance}
\begin{figure}[!htb]
  \begin{center}
  \includegraphics[height=8cm]{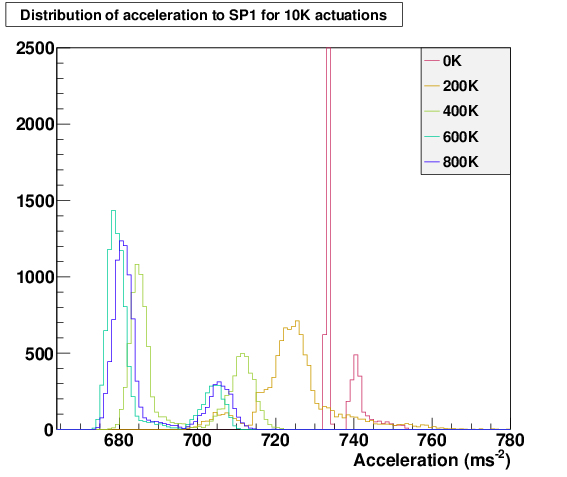}
    \caption{Acceleration recorded for samples of 10K actuations every 200K cycles. 
      The initial drop in acceleration caused by increased bearing wear is apparent. }
    \label{Fig:Performance:AccelerationHist}
  \end{center}
\end{figure}

The bearing performance is monitored using two key parameters.
The first 
 is the acceleration of the shaft, since any increase in friction will also 
 cause a decrease in acceleration. This is calculated from the start
 of the actuation to the first switch-point (see figure~\ref{Fig:Controls:4Quad}), since the
 velocity of the shaft at this point ensures a good time-resolution.
 Unfortunately, any change in the voltage or temperature of the coils also
 has an effect on the coil current and therefore on the acceleration. To help
 reduce these effects the capacitors are charged to a fixed voltage (115\,V) and
 the temperature of the coils is recorded. Figure~\ref{Fig:Performance:AccelerationHist}
 shows the drop in acceleration over time as the bearings wear. 
 
\begin{figure}[!htb]
  \begin{center}
  \includegraphics[height=8cm]{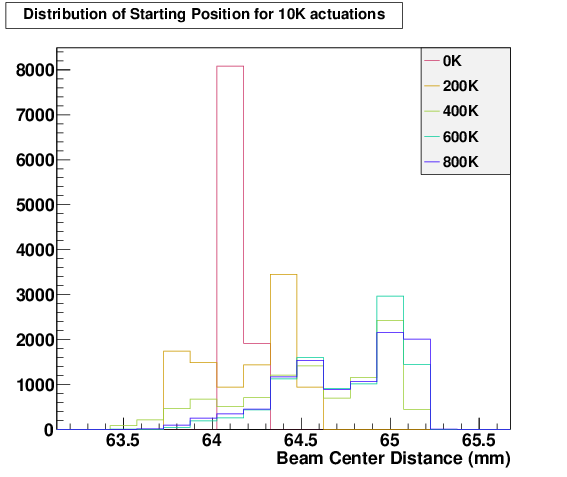}
    \caption{Starting position for samples of 10K actuations every 200K cycles.
      This shows a broadening as the bearings wear. The distribution increases
      unevenly due to a combination of effects from the magnetic fields within the stator 
      and internal feedback used to keep the capture process stable.}
    \label{Fig:Performance:StartingHist}
  \end{center}
\end{figure}

 In addition to decreasing acceleration, the greater friction also allows the
 shaft and magnets to be captured and held further from the zero-force point
 at the centre of the magnetic potential (see section \ref{Sect:PowerElectronics:ZeroForce}). 
 This can be observed at the beginning of the next actuation when the shaft
 position is displaced from its normal starting location. This effect is
 monitored over time by plotting the distribution of starting positions
 for several hundred actuations and calculating the full width. Figure~\ref{Fig:Performance:StartingHist}
 shows the increase in the full width of the starting position. The increasing
 jitter on the starting position also causes variation in the BCD
 due to the symmetry about set-point 1. An increase in jitter in
 BCD will increase the variability of the generated beam-loss
 and hence makes it more likely that the this value exceeds the
 amount permitted by ISIS for that run. 

\subsubsection{Test Programme}
To aid in understand bearing wear, develop monitoring techniques and improve
 bearing life an extensive test programme has been undertaken. The nomenclature
 used to identify the target version is T$n.m$, where $n$ corresponds to the stator
 number and $m$ corresponds to the shaft/bearing combination. For testing, stator
 2 was used, as stator 1 was installed inside ISIS. Each test used a freshly-machined 
 set of VESPEL bearings. For maximum wear, the largest strike of 45\,mm
 (corresponding to 19\,mm BCD) and a high dip rate (0.83\,Hz) were used. 
 For comparison, operation in ISIS
 typically runs at rate of 0.4\,Hz and with a strike of 36\,mm.

Table~\ref{Tab:BearingTests} shows a summary of the bearing tests on stator 2 since the
 development of the FPGA-based control system. The first system tested
 was T2.4, which had a fully quality-assured DLC-coated shaft and a set of
 VESPEL bearings. The test ran for approximately 3 weeks and was disassembled
 at weekly intervals to check for dust. The test was terminated at 1 million
 actuations when the shaft began to lock up due to wear in the anti-rotation
 component. This was resolved by cutting semi-circular reliefs out of the
 bearing as described in section 4.2.3, which was implemented in the subsequent
 test T2.5.

\begin{table}[hbtp]
  \begin{minipage}{\textwidth}
  \begin{center}
    \caption{Summary of bearing tests, ordered chronologically.}
    \label{Tab:BearingTests}
    \begin{tabular}{| c | c | c | c | c |}
    \hline  
    Test & Actuations (K) & Run  & Acceptable & Comments \\
      & & mode & for ISIS? &  \\
    \hline \hline
    T2.4 & 1,000 & WI\footnote{Weekly Inspection. The target was disassembled on a weekly basis to 
      monitor the dust produced by the bearing wear.} & Y & Well-polished shaft, occasional sticking \\
    & & & & during capture (outside ISIS). \\
    \hline
    T2.5 & 4,000 & WI & Y & Weekly inspections caused a \\
    & & & & noticeable disturbance to performance. \\
    \hline
    T2.6 & 1,100 & DS\footnote{Daily Stop. Once a day the mechanism is stopped for an hour to let the system cool and thermally contract.} & Y & Tight bearing-clearances. Sticking during \\
    & & & &    capture noted at end of run. \\
    \hline
    T2.7 & 1,300 & DS & Y & Clearances re-matched to T2.5, minor \\
    & & & & increase in time before sticking during capture. \\
    \hline
    T2.8a & 1,000 & DS & Y &Further increase to bearing-clearance, sticking \\
    & & & &  began near 1 million actuations. Test paused \\
    & & & &  for controller update to help alleviate sticking. \\
    \hline
    T2.8b & 1,000 + 1,500 & DS & Y & T2.8 test resumed with an updated controller \\
    & & & & and enabled a further 1.5 million actuations. \\
    \hline
    
    \end{tabular}
  \end{center}
  \end{minipage}
\end{table}

\begin{figure}[!htb]
  \begin{center}
  \includegraphics[height=8cm]{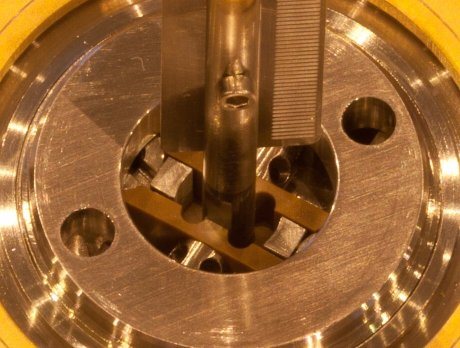}
    \caption{Image of the upper bearing of T2.4 during the final inspection after 4
    million actuations. A small but noticeable amount of dust can be seen on the bearing.}
    \label{Fig:Performance:T24DustPhoto}
  \end{center}
\end{figure}

The next test, T2.5, was set up identically to T2.4 and ran for over 8 weeks and
 4 million actuations. Each week the bearings were inspected for dust and a small,
 but noticeable, amount built up over time 
 (see figure~\ref{Fig:Performance:T24DustPhoto}). In the monitoring
 of acceleration and starting position, a performance increase was noticed after 
 the disturbance caused by each
 inspection. Such disturbance is not possible in ISIS and the weekly inspections
 were abandoned in future tests.

\begin{figure}[!htb]
  \begin{center}
  \includegraphics[height=8cm]{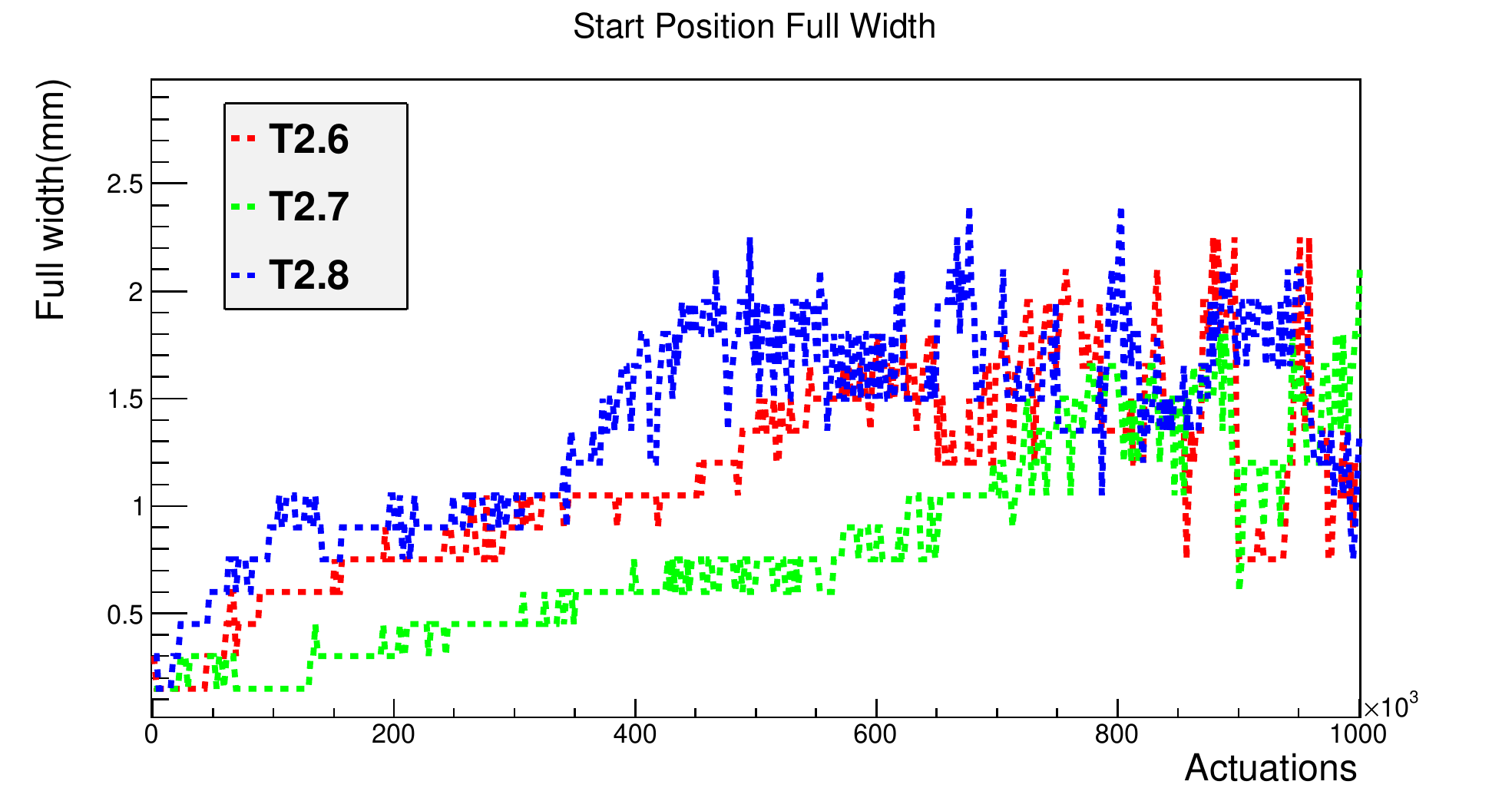}
    \caption{Width of starting position over first 1M actuations for T2.6,7,8. Capture corrections
      (see text) started occurring between 0.9 and 1.0 million actuations.}
    \label{Fig:Performance:MultiStartingHist}
  \end{center}
\end{figure}

The tests of T2.6, T2.7 and T2.8 all ran without intervention and with a daily
 stop of one hour to allow the target to cool and more closely mimic operation in ISIS.
 Each test had slightly different bearing clearances, but all exhibited similar
 performance. The full width of the starting position is shown in 
 figure~\ref{Fig:Performance:MultiStartingHist}, which can
 be seen to broaden steadily over time with T2.7 broadening slowest. All the tests
 ran reliably up to 1 million actuations, but after this the capture of the shaft began to be
 outside the defined limits. The controller is normally able to apply a
 low-force correction to move the target into the correct location before beginning
 the next actuation. However, the increased friction in the bearings caused the shaft to become
 stuck in this state, so an improved control algorithm was put in place for T2.8b. As
 the update was merely to the controller, the target was not disturbed during
 the upgrade procedure and testing was resumed after the controller update. The improved
 algorithm provided a larger force during capture and allowed T2.8b to perform a further 1.5
 million actuations without inspection, bringing the total to 2.5 million actuations.

In the final inspection of the tests T2.4 through to T2.8 there was very little dust
 observed inside the drive, demonstrating the suitability of VESPEL
 bearings for use on ISIS. This enabled the DLC-on-VESPEL design to be approved for use in
 ISIS and T2.9 was used to replace T1.0, the target previously in operation on the accelerator. 

%
\section{Summary}
\label{Sect:Summary}
%
%
A mechanism has been presented which accurately inserts a small target into the halo of the ISIS
proton beam, to generate particles for the Muon Ionisation Cooling 
Experiment\cite{MICE},\cite{MICE_Beam}.
The heart of the device is a linear motor consisting of an array of radial permanent magnets
inside a water-cooled bank of flat coils.
The magnets are mounted on a titanium shaft, the tip of which forms the target.
With appropriate remote position sensing and control of currents through the coils, 
accelerations of over 780\,m\,s$^{-2}$ are achieved during actuations, while the target
remains magnetically levitated between insertions.
Isolation mechanisms are implemented so that, in the case of a fault, the drive can be separated
from the synchrotron both mechanically and from its vacuum system.

The most challenging part of the design has proved to be the sliding bearings which
constrain the motion of the shaft.
The solution adopted uses a diamond-like carbon coating on the shaft engaging with polyimide
inserts.
The target drive has been tested both outside and inside the synchrotron, and detailed
performance data have been recorded.
Reliable operation has been demonstrated for several millions of actuations.
Beam loss caused by the target has been monitored, and the particles produced have enabled the 
operation of the MICE experiment.

%
%
\section*{Acknowledgements}

We gratefully acknowledge the ISIS Division at the STFC Rutherford
Appleton Laboratory for the warm spirit of collaboration  and for
providing access to laboratory space, facilities, and invaluable
support. 
We are indebted to the MICE collaboration, which has provided the
motivation for, and the context within which, the work reported here was
carried out.
We thank Dr.\,N.\,Schofield of the School of Electrical and Electronic Engineering, 
University of Manchester (and formerly of University of Sheffield)
for his design of the original linear motor, 
Dr.\,W.\,Lau of Oxford University for studies of vibrational modes of the target and
Emily Longhi of STFC/Diamond for initial magnetic field mapping.
We would like to acknowledge the work of TecVac of Cambridge in the 
development of diamond-like carbon coatings of the shaft and prototype bearings,
Excel Precision of Gloucester for accurate spark erosion of the lower-shaft 
bore and the apertures in the vane, and 
Multigrind Services Ltd of Rickmansworth for precise grinding of the upper shafts.

This work was supported by the Science and Technology Facilities Council
under grant numbers
PP/E003214/1, PP/E000479/1, PP/E000509/1, PP/E000444/1, and through SLAs
with STFC-supported laboratories.

%
\clearpage
\bibliographystyle{JHEP}

\bibliography{TargetPaper}
%
\end{document}